\documentclass[]{aastex631}

\received{October 29, 2022}
\accepted{December 29, 2022}
\submitjournal{Publications of the Astronomical Society of the Pacific}

\usepackage{graphicx}
\usepackage[caption=false]{subfig}

\shorttitle{Condor Motivation, Configuration, and Performance}
\shortauthors{Lanzetta, Gromoll, Shara, et al.}

\graphicspath{{./}{figures/}}

\begin{document}

\defcitealias{abr2014}{AvD}

\title{Introducing the Condor Array Telescope. 1. Motivation, Configuration,
and Performance}

\author{Kenneth M.\ Lanzetta}
\affiliation{Department of Physics and Astronomy,
Stony Brook University,
Stony Brook, NY 11794-3800, USA}

\author{Stefan Gromoll}
\affiliation{Amazon Web Services,
410 Terry Ave.\ N,
Seattle, WA 98109, USA}

\author{Michael M.\ Shara}
\affiliation{Department of Astrophysics,
American Museum of Natural History,
Central Park West at 79th St.,
New York, NY 10024-5192, USA}

\author{Stephen Berg}
\affiliation{Department of Physics and Astronomy,
Stony Brook University,
Stony Brook, NY 11794-3800, USA}

\author{David Valls-Gabaud}
\affiliation{Observatoire de Paris,
LERMA, CNRS,
61 Avenue de l'Observatoire,
75014 Paris, FRANCE}

\author{Frederick M.\ Walter}
\affiliation{Department of Physics and Astronomy,
Stony Brook University,
Stony Brook, NY 11794-3800, USA}

\author{John K. Webb}
\affiliation{Clare Hall,
University of Cambridge,
Herschel Road,
Cambridge CB3 9AL, UNITED KINGDOM}

\begin{abstract}
The ``Condor Array Telescope'' or ``Condor'' is a high-performance ``array
telescope'' comprised of six apochromatic refracting telescopes of objective
diameter 180 mm, each equipped with a large-format, very low-read-noise
($\approx 1.2$ e$^-$), very rapid-read-time ($< 1$ s) CMOS camera.  Condor is
located at a very dark astronomical site in the southwest corner of New Mexico,
at the Dark Sky New Mexico observatory near Animas, roughly midway between (and
more than 150 km from either) Tucson and El Paso.  Condor enjoys a wide
field of view ($2.29 \times 1.53$ deg$^2$ or 3.50 deg$^2$), is optimized for
measuring {\em both} point sources {\em and} extended, very
low-surface-brightness features, and for broad-band images can operate at a
cadence of 60 s (or even less) while remaining sky-noise limited with a duty
cycle near 100\%.  In its normal mode of operation, Condor obtains broad-band
exposures of exposure time 60 s over dwell times spanning dozens or hundreds of
hours.  In this way, Condor builds up deep, sensitive images while
simultaneously monitoring tens or hundreds of thousands of point sources per
field at a cadence of 60 s.  Condor is also equipped with diffraction gratings
and with a set of He~II 468.6 nm, [O~III] 500.7 nm, He~I 587.5 nm,
H$\alpha$ 656.3 nm, [N~II] 658.4 nm, and [S~II] 671.6 nm narrow-band filters,
allowing it to address a variety of broad- and narrow-band science issues.
Given its unique capabilities, Condor can access regions of ``astronomical
discovery space'' that have never before been studied.  Here we introduce
Condor and describe various aspects of its performance.
\end{abstract}


\section{Introduction}

In the hierarchical or ``bottom-up'' scenario of galaxy formation, galaxies are
formed as gas cools and condenses into dark-matter halos, which are themselves
formed via hierarchical merging and accretion of sub-galactic fragments.
Numerical simulations of structure formation in a dark-energy-dominated,
cold-dark-matter ($\Lambda$CDM) Universe predict that the virialized regions
surrounding massive galaxies should be filled with such fragments, which should
leave signs of their merging and accretion histories in the form of tidal
tails, streams, and debris.  Yet deep images of nearby galaxies fail to detect
satellite galaxies or their tidal effects at anywhere near the density of the
fragments predicted by the simulations.  This discrepancy is known as the
``substructure'' or ``missing satellites'' or ``missing outskirts'' problem
\citep[e.g.][]{kly1999, mer2020}.

A possible solution to the problem is that the satellite galaxies and their
tidal effects {\em are} there, but at a surface brightness below the
sensitivity of deep images obtained by reflecting telescopes.  The
surface-brightness sensitivity of a reflecting telescope is ultimately limited
by systematic effects produced by light that is diffracted by the secondary
mirror and then reflected and scattered by the optical assembly.  Perhaps
reflecting telescopes are simply unsuitable for detecting the very
low-surface-brightness light of satellite galaxies and their tidal effects?
(Refracting telescopes are limited by different systematic effects related to
scattering within and between the lenses.)

This possibility motivated \citet[hereafter AvD]{abr2014} to pioneer a new type
of astronomical telescope optimized for detecting extended, very
low-surface-brightness features.  Specifically, \citetalias{abr2014} combined
eight off-the-shelf Canon telephoto lenses of focal length 400 mm and focal
ratio $f/2.8$ (hence objective diameter 143 mm) with eight off-the-shelf Santa
Barbara Imaging Group STF-8300M CCD cameras and a robotic telescope mount to
form an eight-element ``telephoto array,'' which they named ``Dragonfly.''
Because the lenses are not subject to the same systematic effects that limit
reflecting telescopes, the array is sensitive to low surface brightnesses.
Dragonfly is located in the Northern Hemisphere, in New Mexico, and has since
been upgraded to 48 lenses.

Dragonfly has been used to identify previously unknown very
low-surface-brightness ``ultra-diffuse'' galaxies in the Coma Cluster, which
have sizes comparable to the size of the Milky Way but contain only $0.1 - 1\%$
of the stars \citep{mer2016b}.  The array has also been used to study the
stellar halo around the spiral galaxy M101 \citep{van2014}, to identify
ultra-diffuse galaxies in the field surrounding M101 \citep{mer2014} and in a
group containing the elliptical galaxy NGC 5485 \citep{mer2016b}, and to study
the stellar halos of various nearby galaxies \citep{mer2016a, gil2022}.

A similar telescope was constructed by \citet{spi2019}, who combined 10
telephoto lenses with 10 CCD cameras to form a 10-element telephoto array,
which they named ``Huntsman'' and which is located in the Southern Hemisphere,
at the Siding Spring Observatory in Australia.

Several years ago, we became interested in the possibilities of astronomical
telescopes optimized for detecting extended, very low-surface-brightness
features, and our subsequent musings on some of the issues involved led us to
note the following:
\begin{itemize}

\item Dragonfly is not well suited to measuring point sources.  The plate scale
of its detectors is 2.8 arcsec pix$^{-1}$, hence point sources are
significantly undersampled under seeing conditions typical of good astronomical
sites (say ${\rm FWHM} \approx 1.2$ arcsec) and so are susceptible to blending
and intrapixel sensitivity variations \citep[e.g.][]{kav2001}.  Further, the
Canon lenses are not designed to be diffraction limited, and indeed
\citetalias{abr2014} measured Strehl ratios of their lenses ranging from 0.2 to
0.8, with an average of only 0.4.  While point sources are not central to the
scientific agenda of \citetalias{abr2014}, this is nevertheless a limitation of
the array.

\item The Canon lenses used by Dragonfly suffer significant vignetting, which
limits possibilities of upgrading to larger-format detectors.  Over the $18
\times 14$ mm$^2$ size of the Dragonfly detectors, \citetalias{abr2014}
measured vignetting at the field edges of 20\%.  But the ``image circle'' (i.e.\
the diameter over which the intensity exceeds 60\% of the central intensity) of
the lenses is only $\approx 25$ mm, and over a $36 \times 24$ mm$^2$ detector
(a popular ``full-frame'' format for commercial detectors), the lenses suffer
vignetting at the field edge of 60\%.  Hence the lenses can just barely
illuminate a full-frame detector and certainly not a larger detector.

\item Due to rapidly developing commercial CMOS technology, very large-format
($\approx 150$ Mpix), very rapid-read-time ($< 1$ s), very low-read-noise
($\approx 1$ e$^-$) CMOS cameras were at the time of our deliberations just
becoming available.  Whereas exposures with CCD detectors must be relatively
long (say $\approx 600$ s) to remain sky-noise limited and to maintain high
duty cycle, exposures with these new CMOS detectors could be much shorter,
offering exciting possibilities of very rapid cadence (say $\approx 60$ s or
even shorter) observations.

\end{itemize}
With these considerations in mind, we began to conceive the prototype element
of an ``array telescope'' that was to be constructed using off-the-shelf,
diffraction-limited refracting telescopes of longer focal length (hence finer
plate scale) than Dragonfly and incorporating the new off-the-shelf CMOS
cameras.  It was also clear to us that the possibilities of very rapid cadence
observations opened up by the new CMOS detectors were similar to those
of other array telescopes designed for time-domain astronomy that are currently
under construction or in advanced planning stages, including the Large-Array
Survey Telescope or LAST \citep{ofe2020} and the Argus Array \citep{law2022}.
So in this sense, these other array telescopes offered other relevant points of
comparison.

In mid summer 2019, we were awarded NSF Advanced Technologies and
Instrumentation (ATI) funding for proposal 1910001, ``Concept Feasibility Study
for a High-Performance Telescope Array.''  The proposal outlined a plan to
combine off-the-shelf refracting telescopes with off-the-shelf CMOS cameras to
construct an array telescope to be located at a very dark astronomical site in
the Southern Hemisphere, at the El Sauce Observatory in the Rio Hurtado Valley
of Chile, about 15 km from Gemini and LSST.  We dubbed the nascent telescope
the ``Condor Array Telescope'' or simply ``Condor,'' referencing the Andean
condor that is a national symbol of Chile and a significant character of the
mythologies of various Andean cultures.

But this funding was awarded just months before the Covid-19 pandemic struck in
early 2020, and by the time we began to assemble components of the telescope,
it had become clear to us that it would not be possible to deploy the telescope
to Chile anytime soon.  Accordingly, in early autumn 2020, we began to search
for an alternate (perhaps temporary) site for Condor in the continental
USA---one that could be reached by car from our base of operations in New York
(as flying was frowned upon near the peak of Covid-19).  We eventually selected
a very dark astronomical site in the southwest corner of New Mexico, at the
Dark Sky New Mexico observatory near Animas, roughly midway between (and more
than 150 km from either) Tucson and El Paso.  Over the period early winter
2020 through early summer 2021, we deployed and commissioned Condor to the Dark
Sky New Mexico observatory.  In its new setting, its name perhaps evokes the
California condor, whose native range as recently as 500 years ago likely
included New Mexico \citep{gan1931}.

 Condor shares an objective of very low-surface-brightness sensitivity with
Dragonfly and Huntsman; but Condor differs from these other telescopes in
several significant ways, in terms of motivation, configuration, and mode of
operation.  Here we introduce Condor and describe various aspects of its
performance.  The organization of this manuscript is as follows:  In \S\ 2 we
describe the motivation, in \S\ 3 we describe the configuration, and in \S\ 4 we
describe the performance.  In \S\ 5 we discuss the mode of operation, which 
plays an important role in setting Condor apart from other telescopes.  We
present a summary and conclusions in \S\ 6.

\section{Motivation}

Telephoto arrays like Dragonfly and Huntsman and array telescopes like Condor
operate by building up deep, sensitive images over time, staring at one spot on
the sky and acquiring many individual exposures over dwell times that might
span dozens or even hundreds of hours.  Subsequent analysis forms deep,
sensitive images by coadding the individual exposures.  For Dragonfly,
individual exposure times are typically 600 s, and dwell times can reach many
dozens of hours or up to around 100 hours.

Our thinking with respect to Condor was that if we were going to stare at one
spot on the sky and acquire many individual exposures over dwell times spanning
dozens or hundreds of hours, then we might as well at the same time monitor
point sources near that spot on the sky at an interesting cadence.  This
immediately led to the key design decisions that (1) Condor should be optimized
for measuring {\em both} point sources {\em and} extended, very
low-surface-brightness features and (2) Condor should operate at a rapid
cadence.  Optimizing for point sources requires diffraction-limited optics and
a finer plate scale and hence longer focal length than Dragonfly or Huntsman;
these requirements (together with the requirement of very
low-surface-brightness sensitivity) dictated apochromatic refracting
telescopes.  Operating at a rapid cadence requires a low detector read noise
and a rapid detector read time; these requirements dictated CMOS detectors.
Hence early on, we knew that Condor must be constructed of apochromatic
refracting telescopes equipped with CMOS cameras.

For Condor to simultaneously monitor point sources at a rapid cadence and build
up deep, sensitive images by acquiring many individual exposures requires that
the individual exposures must be sky-noise limited.  This obviously involves
some interplay between cadence, telescope aperture, filter bandpass, plate
scale, and detector read noise.  We explored different combinations of
telescopes, filters, and detectors, considering telescope objective diameters
in the range 150 through 200 mm.  Cost was also a significant driver here, as a
given light-collecting ability might be achieved by using more smaller
telescopes (which implies more cameras, filters, and other peripherals and a
correspondingly higher data rate) or fewer larger telescopes (which implies
fewer cameras, filters, and other peripherals and a correspondingly lower
data rate).  We identified a sweet spot at an objective diameter of 180 mm,
where for a given light-collecting ability the combined cost of the telescopes
plus cameras, filters, and other expensive peripherals was minimized.

There is also an interplay between cadence and the depth of each individual
exposure, with an obvious trade off between shorter, shallower individual
exposures (with a correspondingly higher data rate) or longer, deeper
individual exposures (with a correspondingly lower data rate).  We eventually
settled on a primary cadence of 60 seconds and a primary filter with a
luminance bandpass.  Given a telescope of objective diameter 180 mm and a
low-read-noise, rapid-read-time CMOS camera, observations obtained under such a
configuration would be sky-noise limited even under very dark conditions and
would operate at a duty cycle near 100\%.  And combining six such telescopes
(which is what would fit within the allotted budget), each individual summed
(luminance-bandpass) exposure would reach a depth of $\approx 21$ mag, which is
well matched to the limiting depth of the Gaia catalog \citep{gai2017,
gai2018}.

Further, our thinking with respect to Condor was that an instrument optimized
for very low-surface-brightness sensitivity could fruitfully be directed at
obtaining narrow- as well as broad-band images.  Remarkably, most of the sky
has {\em never} been imaged in the prominent emission lines of astrophysics.
The nearly-completed WHAM Sky Survey \citep[e.g.][]{haf2011} used a Fabry-Perot
spectrometer to scan the entire sky in H$\alpha$ 656.3, but at an angular
resolution of only 1 deg, while the IPHAS and VPHAS+ surveys \citep{wri2016}
achieved higher resolution but covered less than 10\% of the sky.  Deep imaging
observations in He II 468.6, [O III] 500.7, and [S II] 671.6 of $\approx 100$
nearby galaxies have been used to detect supernova remnants, Wolf-Rayet stars,
and planetary nebulae, respectively, but together these observations cover just
a few percent of the sky.  And the state-of-the-art SIGNALS survey
\citep{rou2019} expects to obtain sensitive observations toward $\approx 100$
nearby galaxies using an imaging Fourier transform spectrograph on the CFHT,
but it too will cover only a tiny portion of the sky.  Because most of the sky
remains mostly unknown in most emission lines, here the potential for the
discovery of new phenomena is high.  This led to another key design decision
that (3) Condor should be equipped with a variety of narrow-band filters tuned
to some of the prominent emission lines of astrophysics.

Our ultimate design of the prototype element of Condor satisfied each of the
three key design decisions and was eventually made manifest in the instrument
deployed to New Mexico.  This design may be summarized as follows:

\begin{itemize}

\item Condor is comprised of six apochromatic refracting telescopes of
objective diameter 180 mm and effective focal length 907 mm, which results in a
much finer plate scale than Dragonfly or Huntsman.  It can obtain nearly
Nyquist-sampled images at a good astronomical site and hence, unlike Dragonfly
or Huntsman, is optimized for measuring {\em both} point sources {\em and}
extended, very low-surface-brightness features.

\item Condor is equipped with large-format ($9576 \times 6388$),
very low-read-noise ($\approx 1.2$ e$^-$), very rapid-read-time ($< 1$
s) CMOS cameras, in contrast to the CCD cameras of Dragonfly and Huntsman.  It
enjoys a wide field of view ($2.29 \times 1.53$ deg$^2$ or 3.50 deg$^2$) and
for broad-band images can operate at a cadence of 60 s (or even less) while
remaining sky-noise limited with a duty cycle of nearly 100\%.

\item Like Dragonfly and Huntsman, Condor is equipped with a set of Sloan $g'$,
$r'$, and $i'$ broad-band filters (one each per telescope).  But Condor is also
equipped with luminance filters and diffraction gratings (one each per
telescope) and a set of He~II 468.6 nm, [O~III] 500.7 nm, He~I 587.5 nm,
H$\alpha$ 656.3 nm, [N~II] 658.4 nm, and [S~II] 671.6 nm narrow-band filters
(one per telescope), allowing it to address a variety of broad- and narrow-band
science issues.

\item Condor is equipped with a direct-drive mount that can slew anywhere on
the sky in under 3.5 s, allowing it to efficiently intersperse calibration
observations between science observations throughout the course of each night.

\end{itemize}
Given its unique capabilities, Condor can access regions of ``astronomical
discovery space'' \citep{har1984} that have never before been studied.  We
describe details of the configuration in the following section.

\section{Configuration}

\subsection{Telescopes}

The prototype element of Condor is comprised of of six Telescope Engineering
Company (TEC) 180 mm-diameter f/7 refracting telescopes of focal length 1260
mm.  The telescopes feature diffraction-limited, apochromatic, oil-spaced,
triplet objective lenses with multi-layer coatings on all surfaces and a
CaF$_2$ middle element.  Each optical assembly sits within a lightweight
aluminum tube coated with an anti-reflection coating, and each tube assembly
contains custom-designed and -fabricated sharp-edged baffles throughout the
interior.  The baffles of two of the telescopes are coated with a standard
black paint, while the baffles of four of the telescopes are coated with
Singularity Black carbon-nanotube paint, thus allowing for the possibility of
assessing the performance of very black carbon-nanotube paint on the telescope
baffles.

\subsection{Focal-Reducing Field Correctors}

Each TEC telescope is equipped with an Astro-Physics (A-P) 0.72x QUADTCC-TEC180
four-element telecompressor corrector.  These correctors when used with the TEC
180 mm-diameter telescopes yield an effective focal length of 907 mm and an
effective focal ratio of f/5.0.  The effective focal ratio of the six
telescopes together, considered as a single telescope, is f/2.0.

\subsection{CMOS Cameras}

Each TEC telescope is equipped with a ZWO ASI6200MM monochrome CMOS camera,
which is based on the large-format Sony IMX455 back-illuminated detector.  The
cameras feature USB3.0 interfaces and 256 MB DDR3 caches.  The cameras use
Peltier coolers to cool the detectors to as much as 40 C below the ambient
temperature.  Specifications of the CMOS camera detectors are summarized in
Table 1.

\begin{table}[ht]
\centering
\hspace{-0.35in}
\begin{tabular}{p{2.5in}c}
\multicolumn{2}{c}{{\bf Table 1:}  CMOS Camera Detector Specifications} \\
\hline
\hline
Detector size \dotfill & $36 \times 24$ mm$^2$ \\
Detector format \dotfill & $9576 \times 6388$ \\
Pixel size \dotfill & 3.76 $\mu$m \\
Maximum full-well capacity \dotfill & 51 ke$^-$ \\
Read noise \dotfill & 1.3 to 3.5 e$^{-}$ \\
ADC bits \dotfill & 16 \\
Peak quantum efficiency \dotfill & 80\% \\
Maximum full-resolution frame rate \dotfill & 3.2 fps \\
\hline
\end{tabular}
\end{table}

\vspace{-0.2in}

\subsection{Filter Wheels}

Each TEC telescope is equipped with a ZWO EFW 7/2'' seven-position filter
wheel.  Each filter wheel is capable of holding seven 2-inch round filters.

\subsection{Filters and Diffraction Gratings}

Each TEC telescope is equipped with (1) a Sloan $g'$ filter, (2) a Sloan $r'$
filter, (3) a Sloan $i'$ filter, (4) a luminance filter, and (5) a narrow-band
filter, each manufactured by Chroma Technology Corp.  The six narrow-band
filters are tuned to emission lines of (1) He II 468.6 nm, (2) [O III] 500.7
nm, (3) He I 587.5 nm, (4) H$\alpha$ 656.3 nm, (5) [N II] 658.4 nm, and (6) [S
II] 671.6 nm.  The six narrow-band filters have widths of $\approx 4$ nm.
Typical response functions of the various filters (as measured by the
manufacturer) are shown in Figure 1.  The peak transmission of each filter
approaches 100\%.  The Sloan $i'$ and luminance filters exhibit infrared leak,
which is more significant for the luminance filters.

\begin{figure}[ht!]
\centering
\subfloat{
  \includegraphics[width=0.33\linewidth]{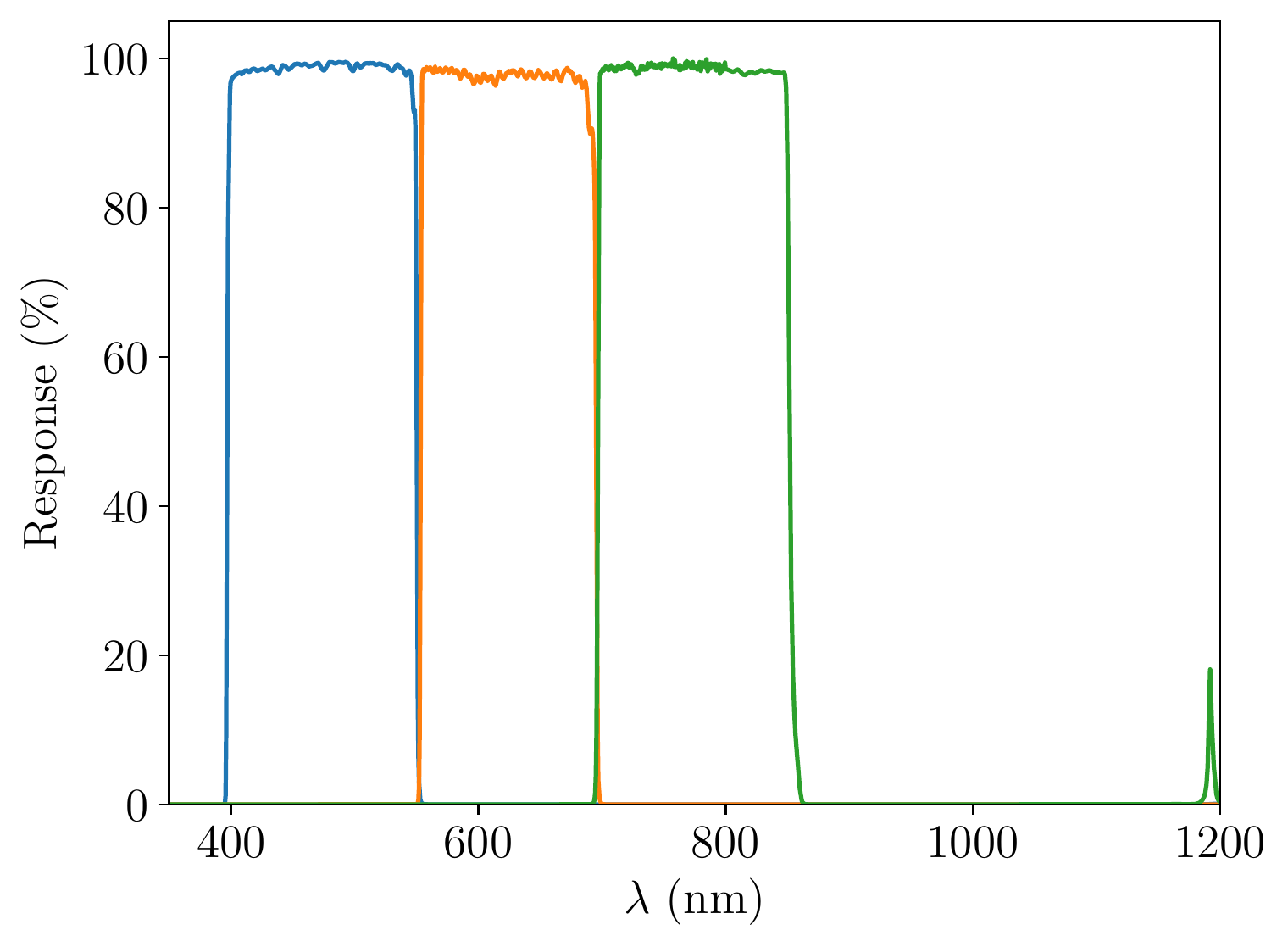}
}
\subfloat{
  \includegraphics[width=0.33\linewidth]{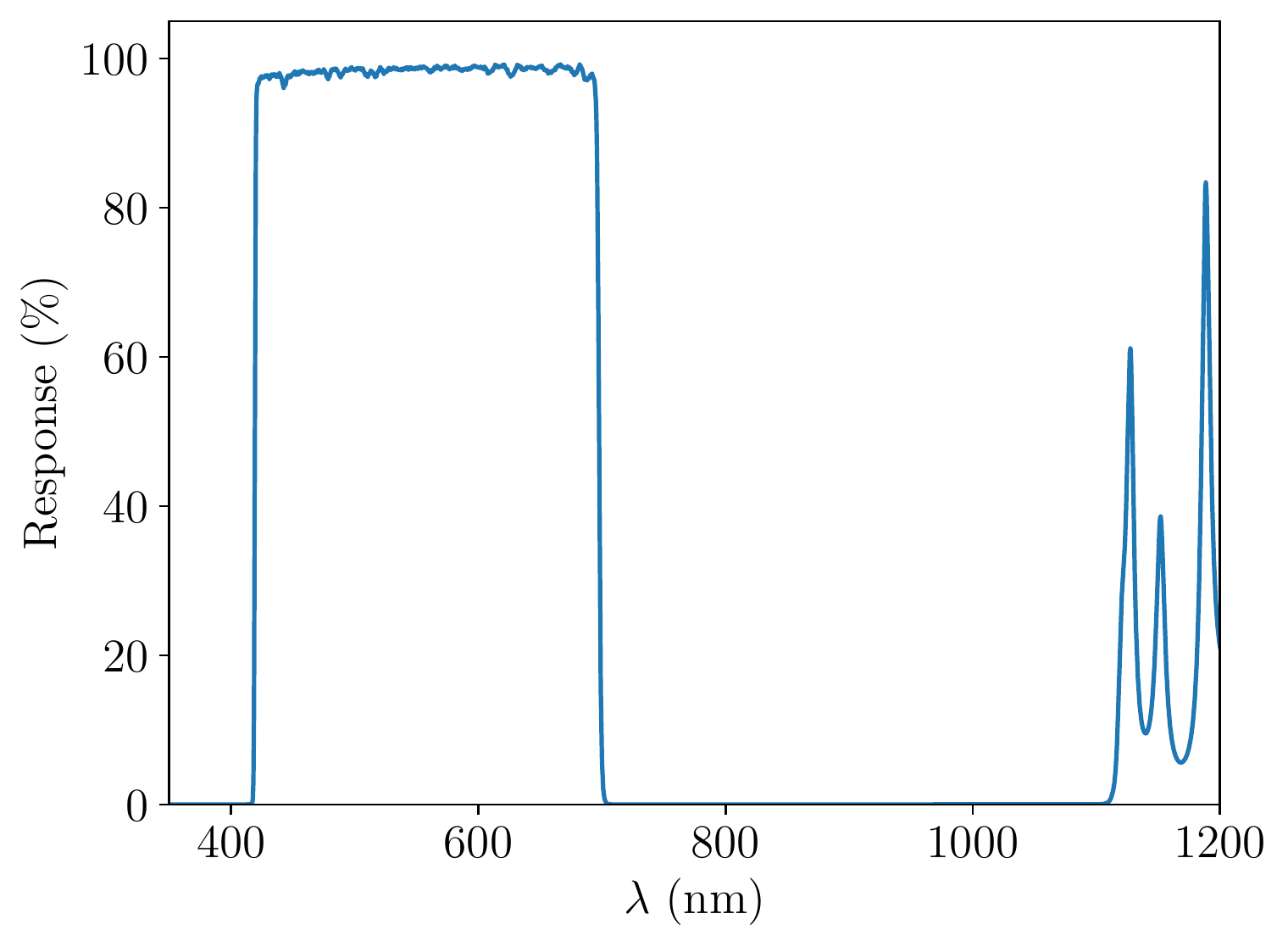}
}
\subfloat{
  \includegraphics[width=0.33\linewidth]{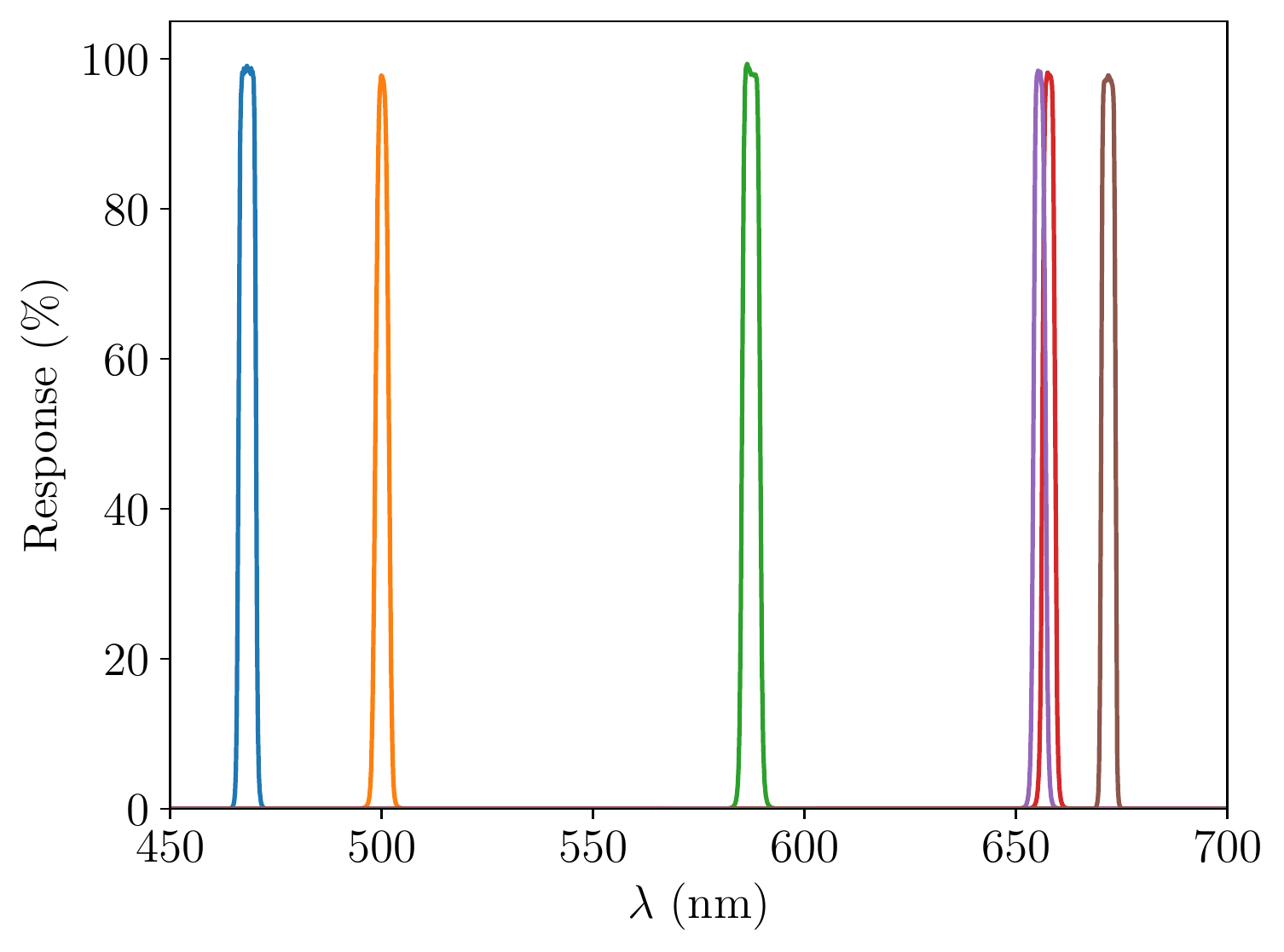}
}
\caption{Typical response functions of the various filters (as measured by
manufacturer):  ({\em a, left}) the Sloan $g'$ (blue), $r'$ (orange), and $i'$
(green) filters,  ({\em b, middle}) the luminance filter, and ({\em c, right})
the He II 468.6 nm (blue), [O III] 500.7 nm (orange), He I 587.5 nm (green),
H$\alpha$ 656.3 nm (purple), [N II] 658.4 nm (red), and [S II] 671.6 nm (brown)
filters.}
\end{figure}

Each TEC telescope is also equipped with a Star Analyzer 200 diffraction
grating distributed by Field Tested Systems, LLC.  The gratings are ruled at 100
lines mm$^{-1}$ and blazed in first-order at a wavelength of 550 nm and provide
a resolution of $R \equiv \lambda/\Delta \lambda \approx 200$.  The dispersion
directions of the various diffraction gratings are oriented in such a way that
they are rotated with respect to one another, which can significantly aid in
untangling the various overlapping spectra of the various sources in the field
of view.

\newpage

\subsection{Focusers}

Each TEC telescope is equipped with an Optec TCF-Leo low-profile motorized
focuser.  The focusers have a step size of 0.08 $\mu$m and a maximum range of
travel of $\approx 9$ mm.

\subsection{Dust Covers}

Each TEC telescope is equipped with an Optec Alnitak motorized dust cover.

\subsection{Mount}

The six TEC telescopes are mounted onto a Planewave L-600 half-fork mount with
equatorial wedge.  The mount features direct-drive motors and precision
encoders on each axis and slew speeds of up to 50 deg s$^{-1}$  and is not
subject to backlash or periodic error.  The telescopes are attached to the
mount using a custom-designed and -fabricated bracket that allows the
telescopes to be independently collimated with respect to each other.

\subsection{Site}

Condor is located at a very dark astronomical site in the southwest corner of
New Mexico, at the Dark Sky New Mexico observatory near Animas, roughly midway
between (and more than 100 miles from either) Tucson and El Paso, at an
elevation of 1341 m.  Details of the site are summarized in Table 2.

\begin{table}[ht]
\centering
\hspace{-0.35in}
\begin{tabular}{p{1.5in}c}
\multicolumn{2}{c}{{\bf Table 2:}  Details of the Site} \\
\hline
\hline
Latitude \dotfill & 31.94638$^\circ$ N \\
Longitude \dotfill & 108.89755$^\circ$ W \\
Elevation \dotfill & 1341 m \\
\hline
\end{tabular}
\end{table}

The telescope is housed in an observatory building with a roll-off roof.
Images of Condor in its observatory building at the Dark Sky New Mexico
observatory are shown in Figure 2.

\begin{figure}[ht!]
\centering
\subfloat{
  \includegraphics[width=0.33\linewidth, angle=-90, origin=c]{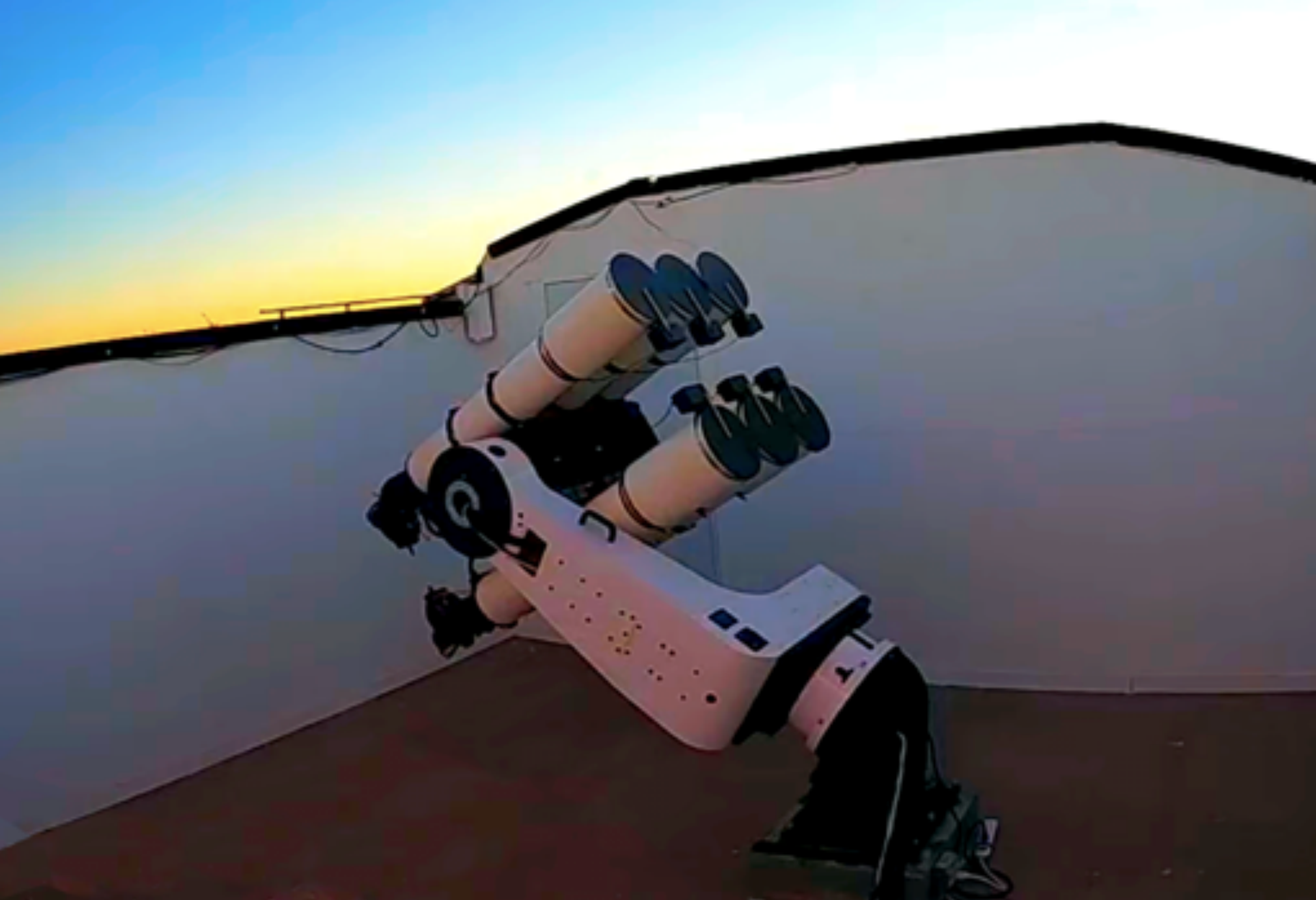}
}
\subfloat{
  \includegraphics[width=0.33\linewidth, angle=-90, origin=c]{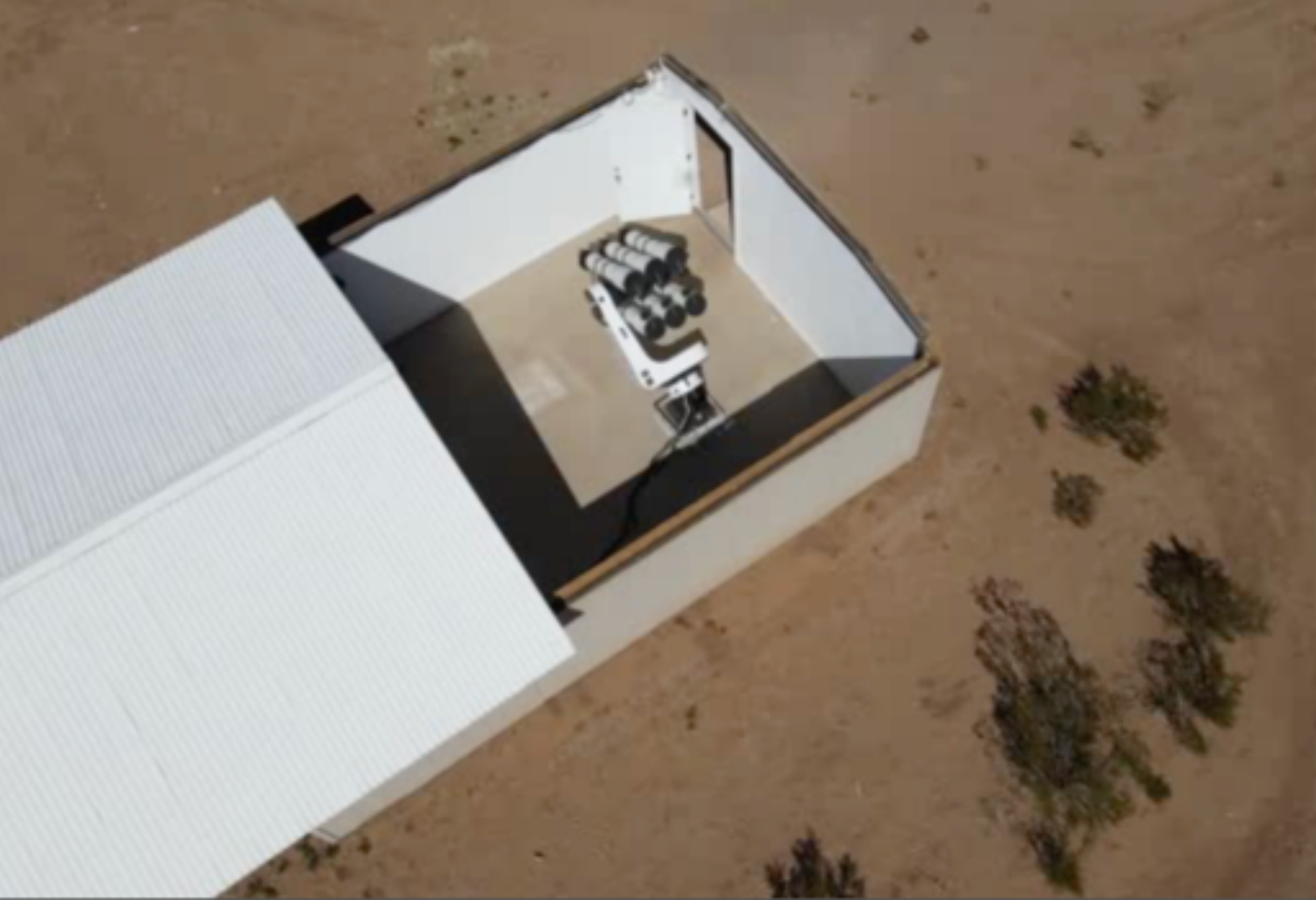}
}
\vspace{-0.5in}
\caption{Images of Condor in its observatory building with a roll-off roof at
the Dark Sky New Mexico observatory near Animas, New Mexico.  In both images,
the dust covers are closed.}
\end{figure}

\subsection{Control and Acquisition Computers}

The CMOS cameras, filter wheels, focusers, dust covers, mount, and observatory
roof are controlled by and data are acquired by eight Raspberry Pi 4 ``control
and acquisition'' computers running Ubuntu 20.04, which are located in a
control room adjacent to the telescope.  Each Raspberry Pi 4 computer is
equipped with a 128 GB SATA III solid-state drive connected via USB3.0, and
each computer and drive assembly is mounted into a custom-designed and
-fabricated 3-D-printed enclosure, and the eight computer and drive enclosures
are mounted into a server rack.  The server rack also contains a router, a 1 TB
solid-state drive, and environmental monitor, an IP power distribution unit, a
power distribution unit, and three focus controller hubs.  The 1 TB solid-state
drive is used to stage data, the environmental monitor is used to monitor
temperature and humidity and the state of the observatory roof, the IP power
distribution unit is used to remotely cycle power to the computers and other
peripherals, and the focus controllers are used to control the focusers.  An
image of the Raspberry Pi 4 computers and the server rack is shown in Figure 3.

\begin{figure}[ht!]
\centering
\includegraphics[width=0.40\linewidth, angle=-90, origin=c]{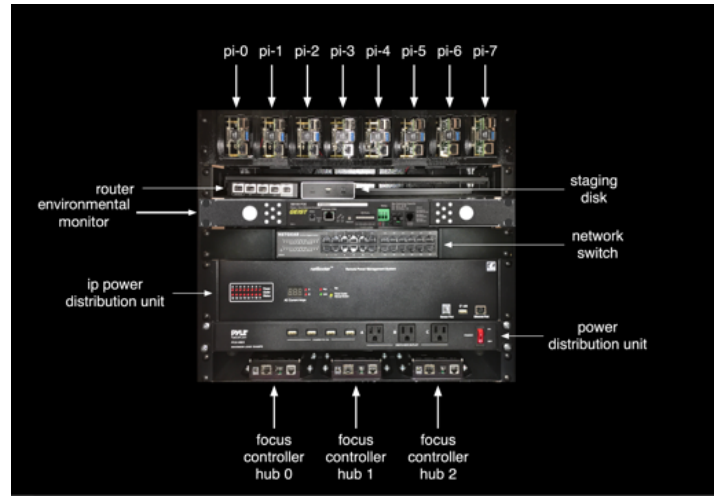}
\vspace{-0.5in}
\caption{Image of Raspberry Pi 4 computers and server rack.  Server rack
contains eight computer and drive enclosures, a router, a 1 TB solid-state
drive, and environmental monitor, an IP power distribution unit, a power
distribution unit, and three focus controller hubs.}
\end{figure}

Each CMOS camera is connected to a Raspberry Pi 4 computer via USB3.0.  The
other peripherals (filter wheels, focusers, dust covers, mount, and observatory
roof) are connected to Raspberry Pi 4 computers via USB2.0 or Ethernet.  In
operation, one Raspberry Pi 4 computer acts as a ``sequencer'' and sends
instructions to other Raspberry Pi 4 computers, which in turn send instructions
to the various peripherals.  As images are acquired, they are temporarily
staged to the 128 GB solid state drives and then promptly transferred to
storage and analysis computers on the campus of Stony Brook University
(described in \S\ 3.11) and from there to storage and analysis computers at the
American Museum of Natural History.  The Raspberry Pi 4 computers continuously
report information (related to status of the observations, status of the
peripherals, status of the computers themselves, and weather) to a database
running on one of the storage and analysis computers; these communications are
cached locally, ensuring that they are not lost in the event of a disruption to
the storage and analysis computers or a loss of Internet connectivity.  Current
and predicted weather is continuously monitored via the Tomorrow.io weather
API; these weather observations are used to assess local conditions bearing on
whether the observatory roof should be opened or closed and are recorded for
later use in science analysis.

\subsection{Storage and Analysis Computers}

Data acquired by Condor are stored on and analyzed by six Dell PowerEdge R720xd
``storage and analysis'' computers running Ubuntu 20.04, which are located at
the Data Center on the campus of Stony Brook University.  Each Dell R720xd
computer is equipped with one or more solid-state drives, which are used to
store the operating system and local scratch space; several high-capacity (14
to 18 TB) hard disk drives, which are used to store data; and 768 GB DDR3
memory.  The various Dell R720xd computers each run MinIO, which is an
open-source object storage system that is functionally compatible with AWS S3,
and data acquired by Condor are ultimately stored under MinIO clusters.  The
various MinIO clusters are configured for a factor two redundancy, and the
system is robust against loss of drives and loss of computers.  One Dell
R720xd computer hosts the Condor web site and runs a PostgreSQL database, and
all six Dell R720xd computers participate in the analysis of data.

\newpage

\section{Performance}

Condor obtained its first-light image on March 7, 2021 and has since acquired
thousands of hours of observations.  Here we describe the performance based on
analysis of some of these observations.

\subsection{Characteristics of the CMOS Detectors}

Here we describe characteristics of the Sony IMX455 detectors used in the ZWO
ASI6200MM monochrome CMOS cameras.

\subsubsection{Nominal Gain, Read Noise, Full-Well Capacity, and Dynamic Range}

Various properties of the CMOS detectors (as specified by the manufacturer) are
shown in Figure 4 as functions of the ``gain setting,'' including nominal
full-well capacity, gain, dynamic range, and read noise.  It is apparent from
Figure 4 that there are two obvious choices of gain setting:  a ``low-gain''
setting of 0 yields a nominal gain of 0.8 e$^-$ ADU$^{-1}$, a nominal full-well
capacity of 51 ke$^-$, a nominal read noise of 3.5 e$^-$, and a dynamic range
of just under 14 stops, while a ``high-gain'' setting of 100 yields a nominal
gain of 0.27 e$^-$ ADU$^{-1}$, a nominal full-well capacity of 19 ke$^-$, a
nominal read noise of 1.5 e$^-$, and a dynamic range of again just under 14
stops.  In practice, we obtain observations at the high-gain setting of 100.

\begin{figure}[ht!]
\centering
\subfloat{
  \includegraphics[width=0.40\linewidth]{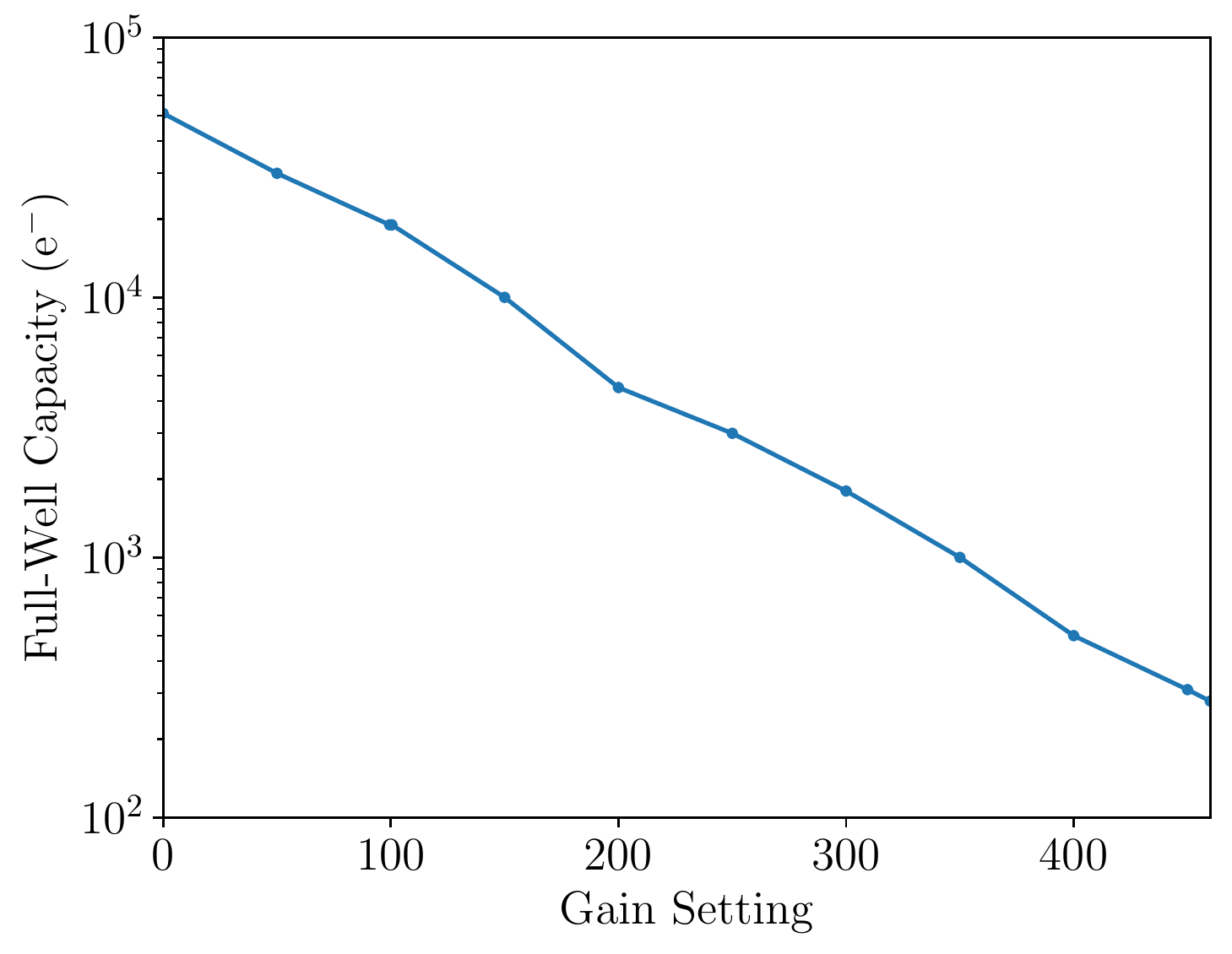}
}
\subfloat{
  \includegraphics[width=0.40\linewidth]{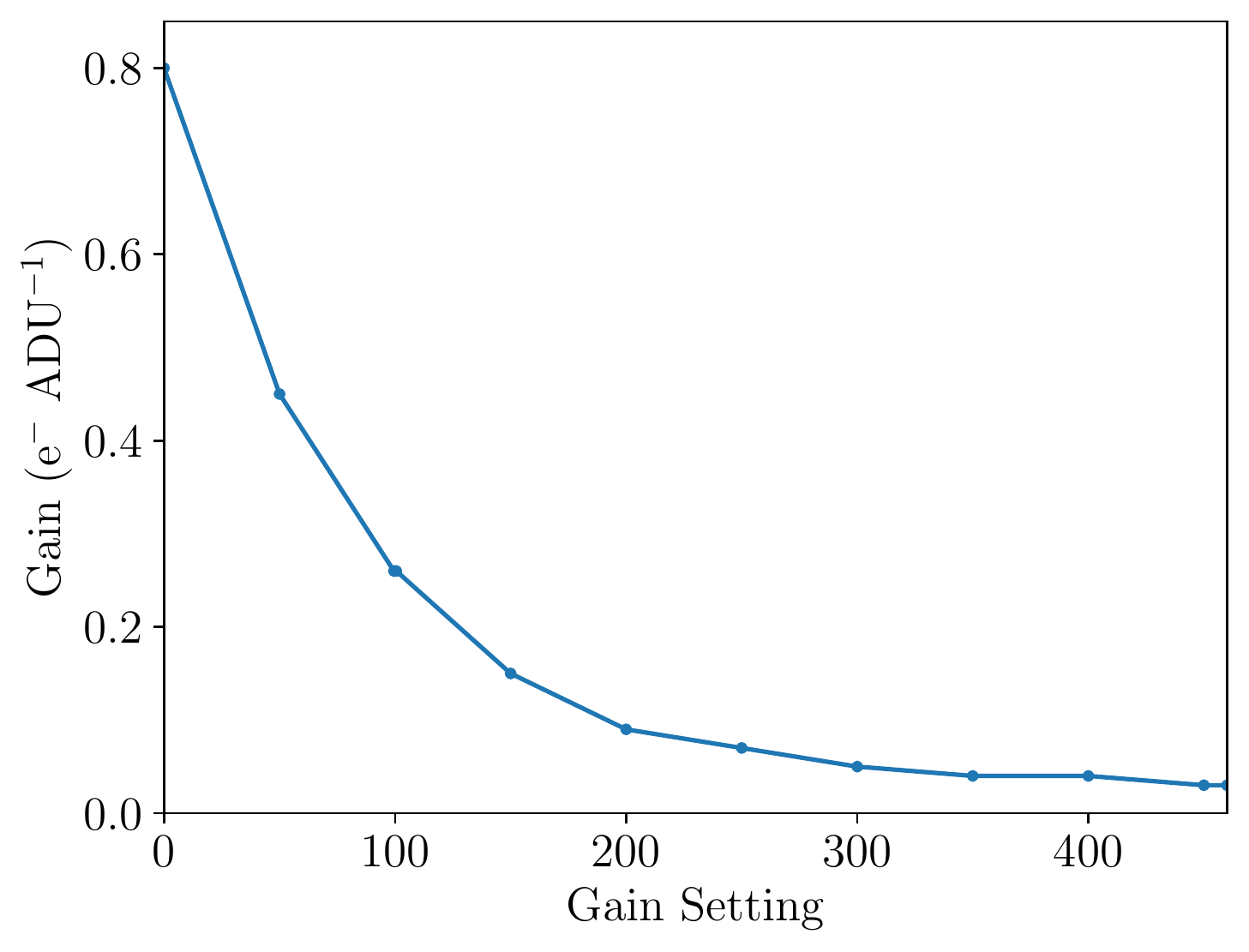}
}
\\
\subfloat{
  \includegraphics[width=0.40\linewidth]{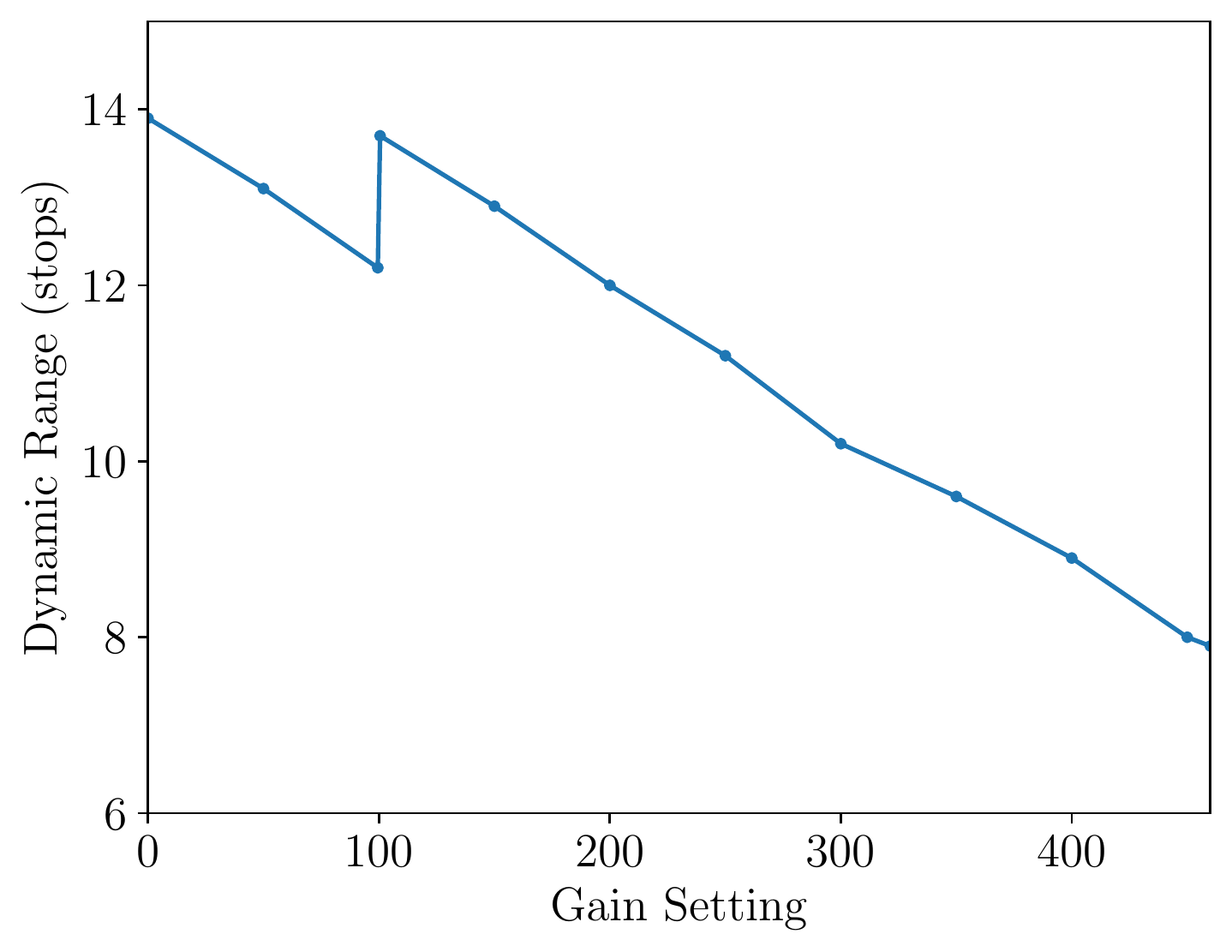}
}
\subfloat{
  \includegraphics[width=0.40\linewidth]{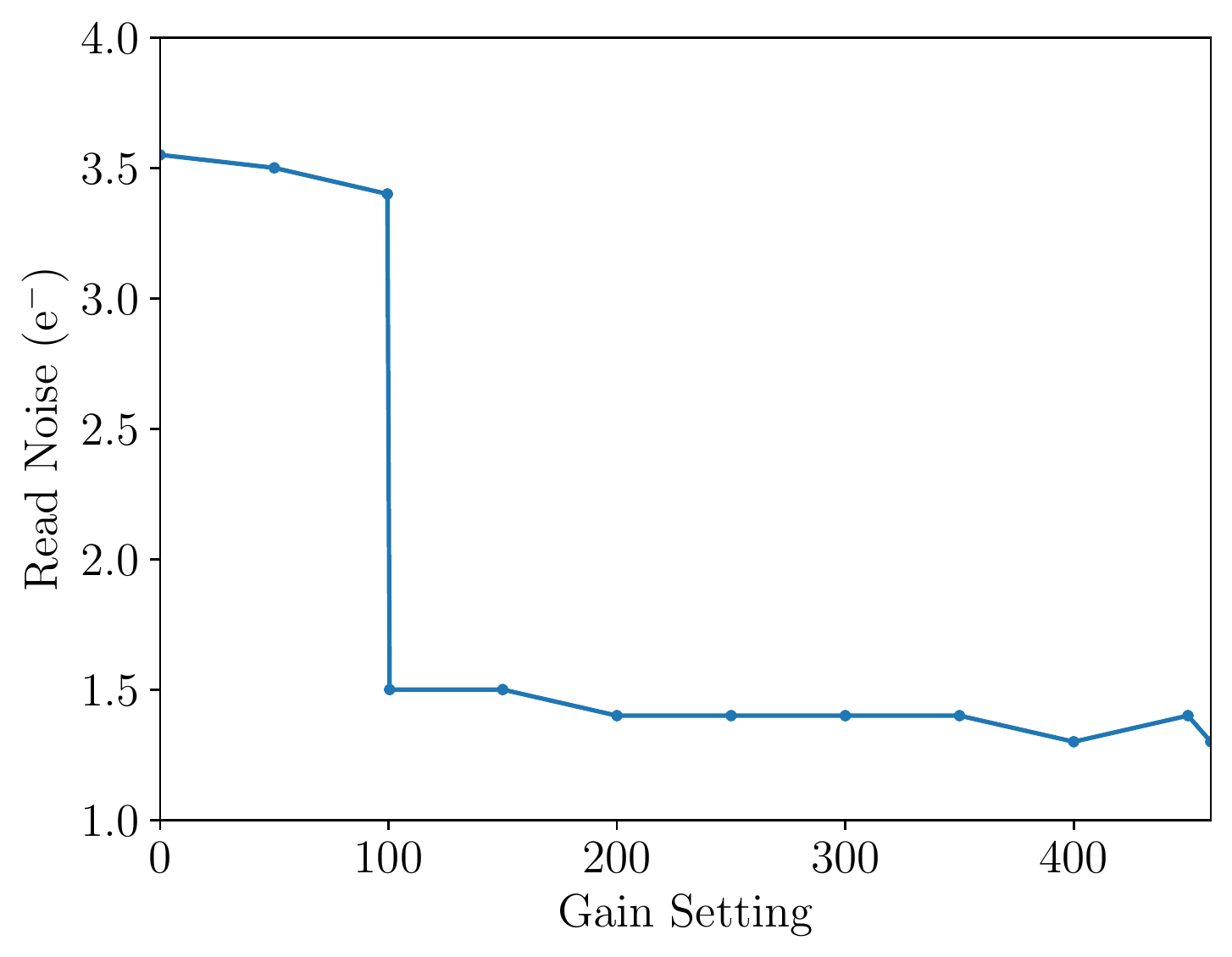}
}
\caption{Various properties of CMOS detectors (as specified by manufacturer):
({\em a, upper left}) full-well capacity versus gain setting, ({\em b, upper
right}) gain versus gain setting,  ({\em c, lower left}) dynamic range versus
gain setting, and ({\em d, lower right}) read noise versus gain setting.  In
practice, we obtain observations at the ``high-gain'' setting of 100.}
\end{figure}

\subsubsection{Bias and Pixel Noise}

Condor obtains frequent (typically daily) sequences of 500 zero-exposure ``bias
images'' of each CMOS detector in order to assess bias and pixel noise.  From
these sequences of bias images, we produce ``master bias images'' that consist
of two planes:  one plane contains the pixel-by-pixel median of the sequence,
and the other plane contains the pixel-by-pixel standard deviation about the
median of the sequence.  We take the median plane to be the pixel-by-pixel bias
level, and we take the standard deviation plane to be the pixel-by-pixel pixel
noise level.  In practice, we obtain observations at a bias setting of 10,
which yields a nominal bias level of 100 ADU.  Adopting the nominal gain 0.27
e$^-$ ADU$^{-1}$ described in \S\ 4.1.1, this corresponds to a nominal bias
level in electron units of 27 e$^-$.

Pixel-by-pixel comparisons of the bias levels of four master bias images
obtained a day apart, a week apart, and nearly six months apart are shown in
Figure 5.  From Figure 5, it apparent that:  (1) Most pixels ($> 99\%$) always
exhibit a bias level very close to 100 ADU.  (2) Some pixels ($\lesssim 1\%$)
exhibit a bias level less than or greater than 100 ADU in a way that is stable
versus time in the sense that the bias levels of these pixels are highly
correlated with near unit slope between various pairs of master bias images.
And (3) a small fraction ($\lesssim 0.002\%$) of pixels exhibit a bias level
very close to 100 ADU after previously exhibiting a bias level less than or
greater than 100 ADU---or vice versa---in such a way that the bias levels of
these pixels are not highly correlated between various pairs of master bias
images.  We find it notable that the fraction of pixels with bias levels that
are not highly correlated---even nearly six months apart---is $\lesssim
0.002\%$.  The various standard deviations between the various pairs of images
of Figure 5 are $\approx 0.47$ ADU.  Similar results are obtained for different
combinations of master bias images (acquired up to and beyond a year apart) and
for master bias images of different detectors.  We conclude that the
pixel-by-pixel bias levels of the CMOS detectors are very stable over long
periods of time.

\begin{figure}[ht!]
\centering
\subfloat{
  \includegraphics[width=0.33\linewidth]{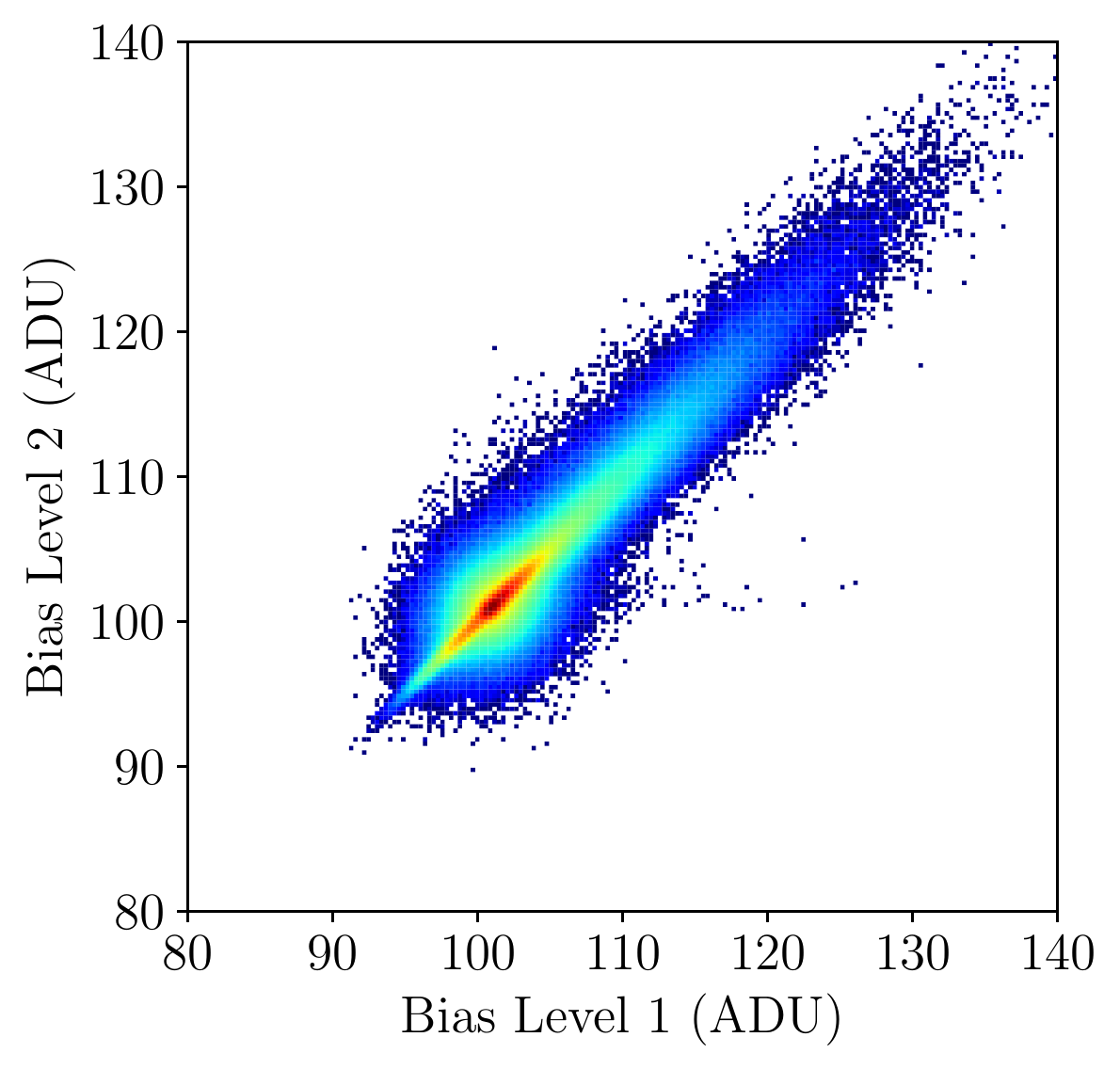}
}
\subfloat{
  \includegraphics[width=0.33\linewidth]{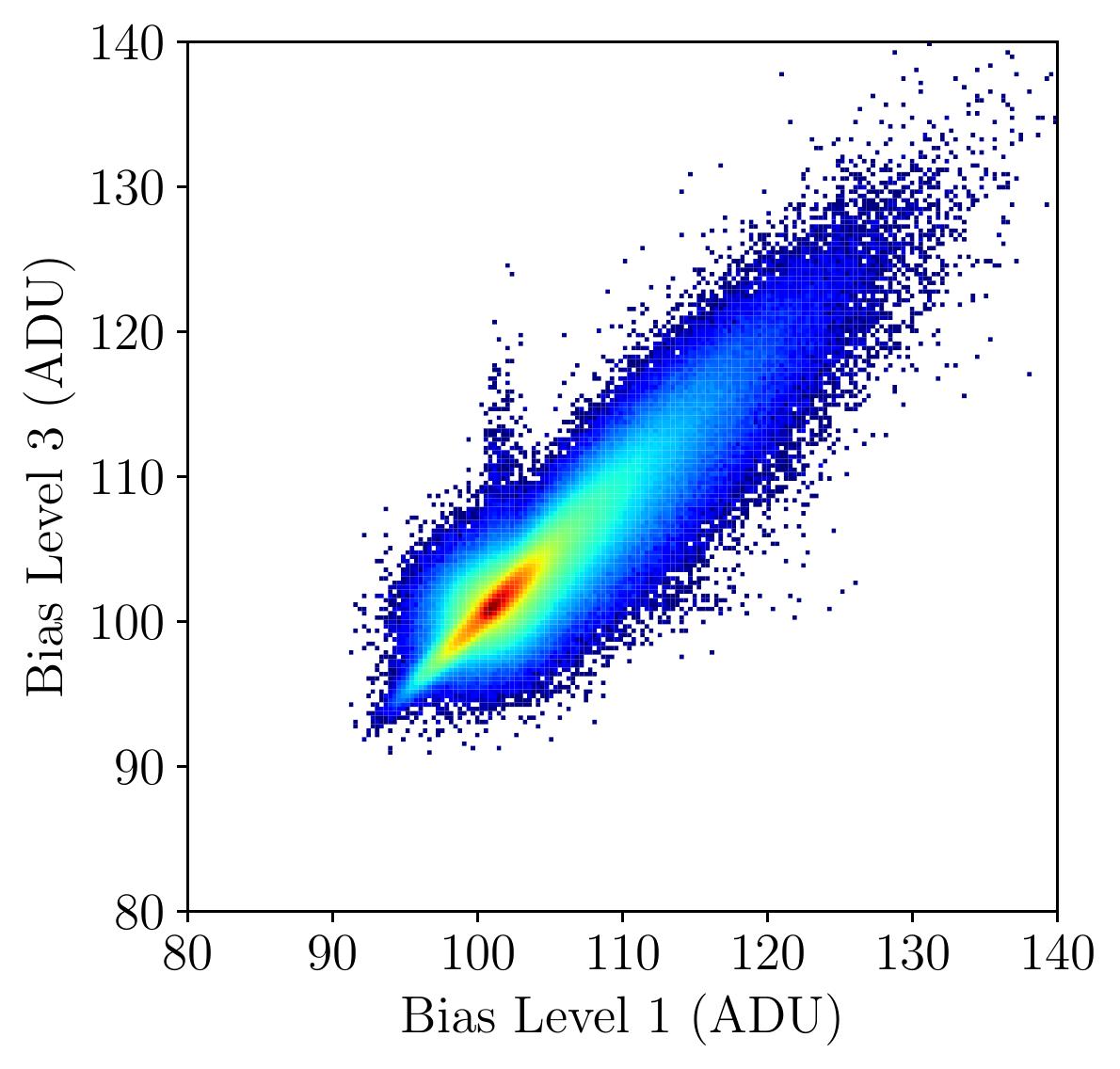}
}
\subfloat{
  \includegraphics[width=0.33\linewidth]{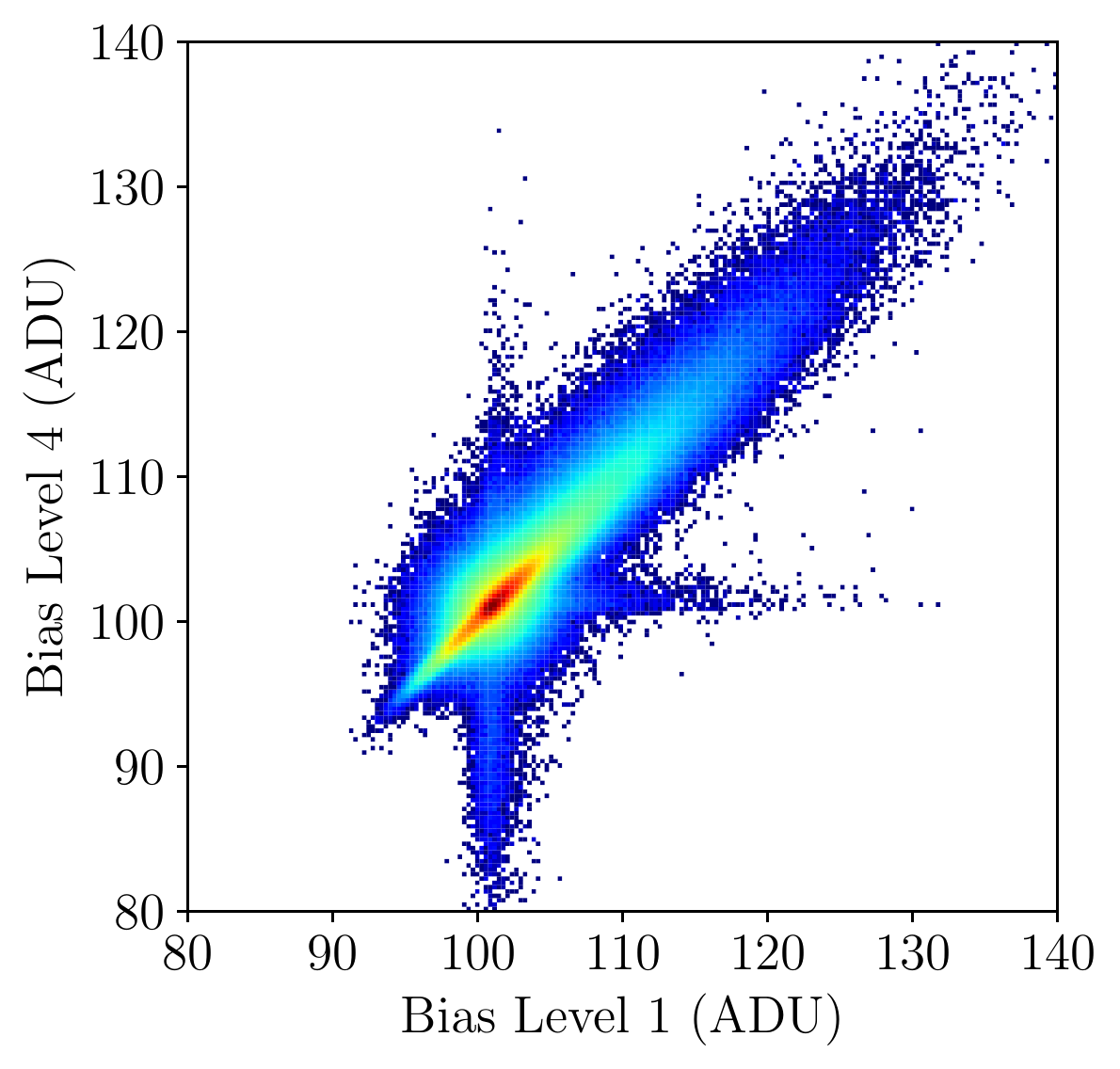}
}
\caption{Pixel-by-pixel comparisons of bias levels of ``master bias'' images 1
(obtained on on May 27, 2022), 2 (obtained on May 26, 2022), 3 (obtained on May
20, 2022), and 4 (obtained on November 22, 2021), all obtained by the same
detector of telescope 4.  (All comparisons are with respect to master bias
image 1.) Displays are logarithmic heat maps, with red representing largest
values and blue representing smallest values.}
\end{figure}

Pixel-by-pixel comparisons of the standard deviations of four master bias
images obtained a day apart, a week apart, and nearly six months apart are
shown in Figure 6.  From Figure 6 it is apparent that:  (1) Most pixels ($>
90\%$) always exhibit a standard deviation very close to 4.3 ADU.  (2) Some
pixels ($\sim 10\%$) exhibit a standard deviation greater than 4.3 ADU in a way
that is stable versus time in the sense that the standard deviations of these
pixels are highly correlated between various pairs of master bias images.  (3)
A small fraction ($\lesssim 0.001\%$) of pixels exhibit a standard deviation
very close to 4.3 ADU after previously exhibiting a standard deviation greater
than 4.3 ADU---or vice versa---in such a way that the standard deviations of
these pixels are not highly correlated between various pairs of master bias
images.  And (4) a small fraction ($\lesssim 0.001\%$) of pixels exhibit a
standard deviation greater than 4.3 ADU after previously exhibiting a standard
deviation close to 4.3 ADU in such a way that the standard deviations of these
pixels are highly correlated between various pairs of master bias images but
with a slope greater than unity.  We find it notable that the fraction of
pixels with standard deviations that are not highly correlated or that are
highly correlated with a slope greater than unity---even nearly six months
apart---is $\lesssim 0.001\%$.  The various standard deviations between the
various pairs of images of Figure 6 are $\approx 0.61$ ADU.  Similar results
are obtained for different combinations of master bias images (acquired up to
and beyond a year apart) and for master bias images of different detectors.  We
conclude that the pixel-by-pixel standard deviations of the CMOS detectors are
very stable over long periods of time.

\begin{figure}[ht!]
\centering
\subfloat{
  \includegraphics[width=0.33\linewidth]{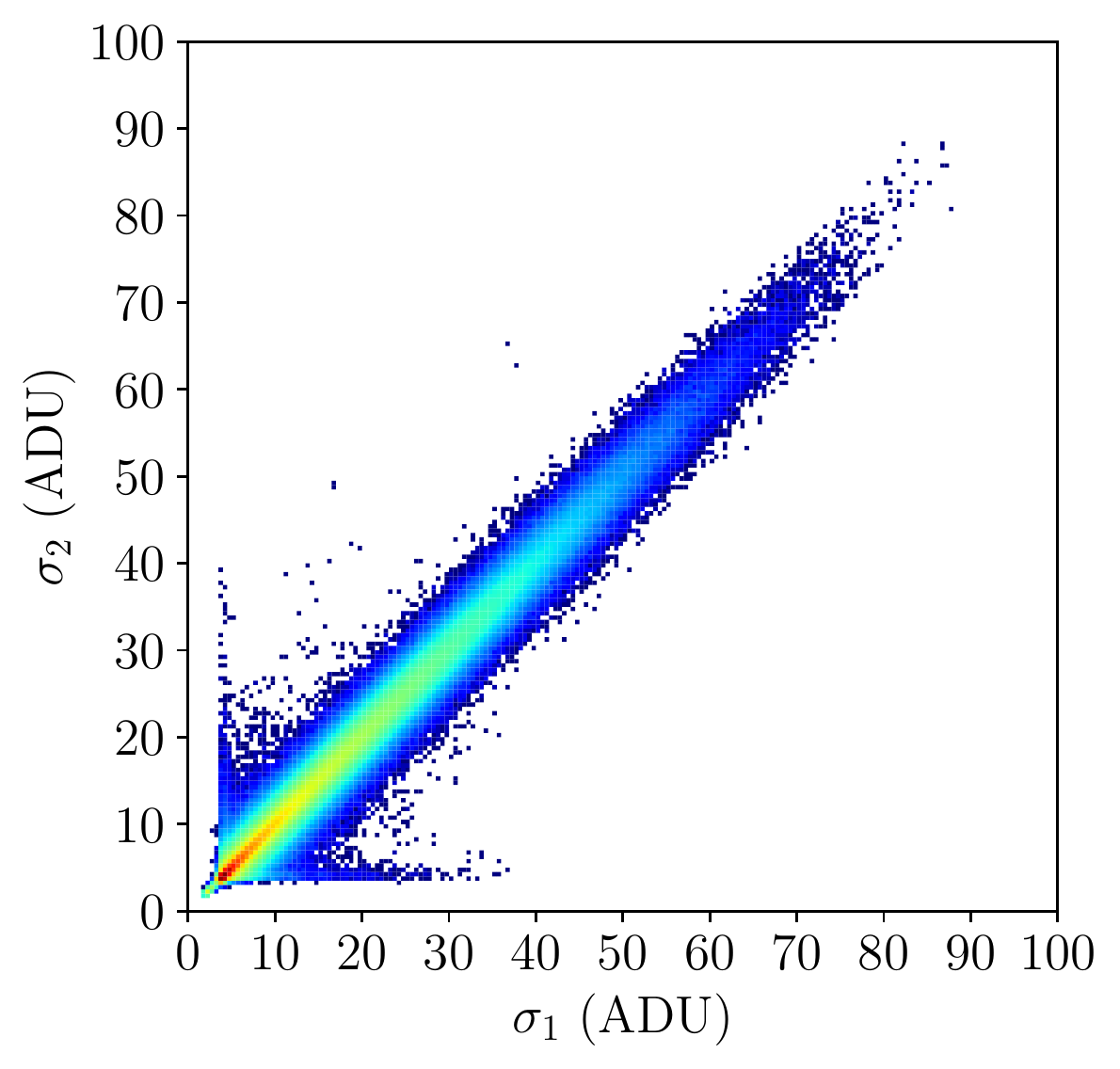}
}
\subfloat{
  \includegraphics[width=0.33\linewidth]{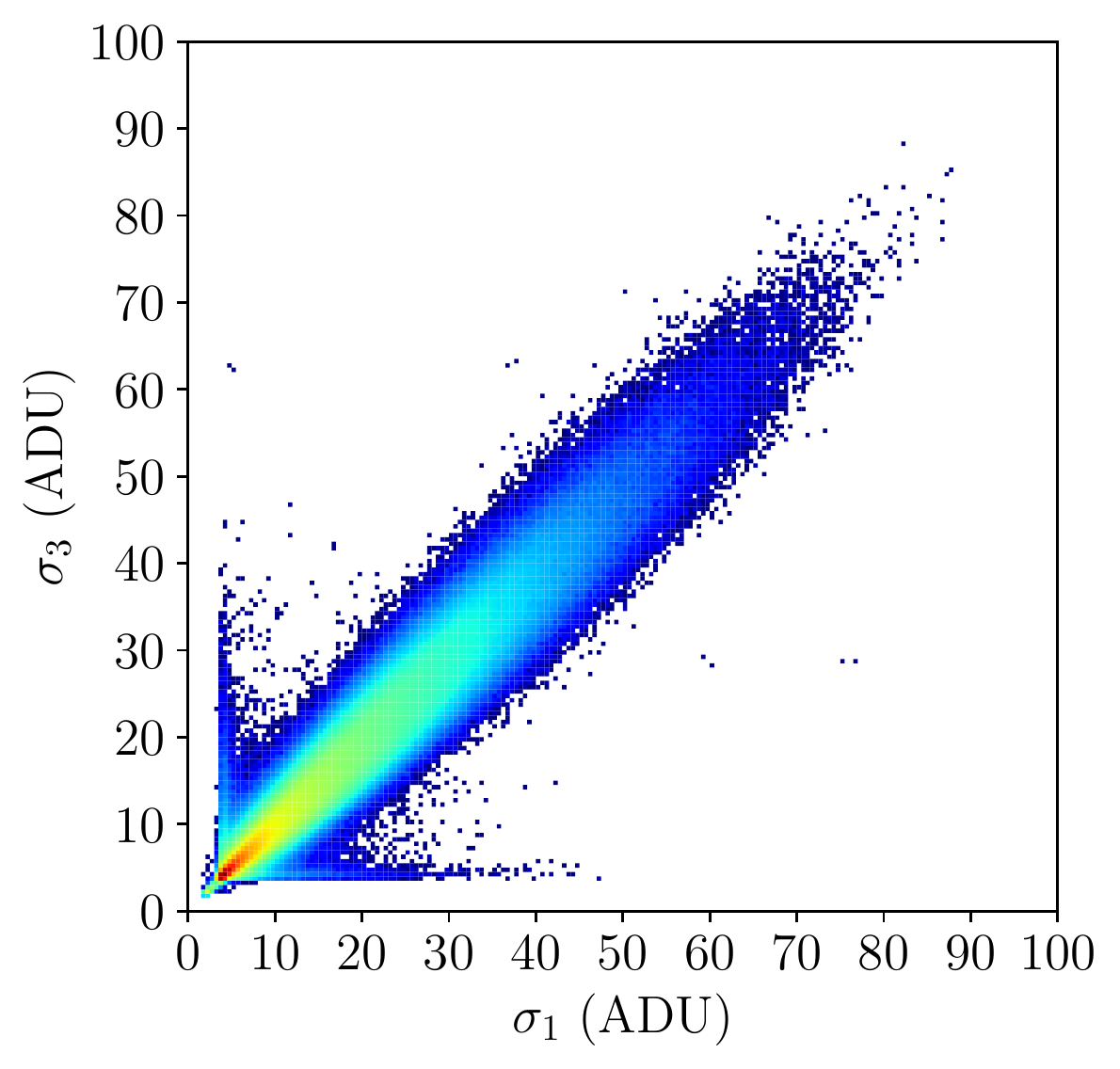}
}
\subfloat{
  \includegraphics[width=0.33\linewidth]{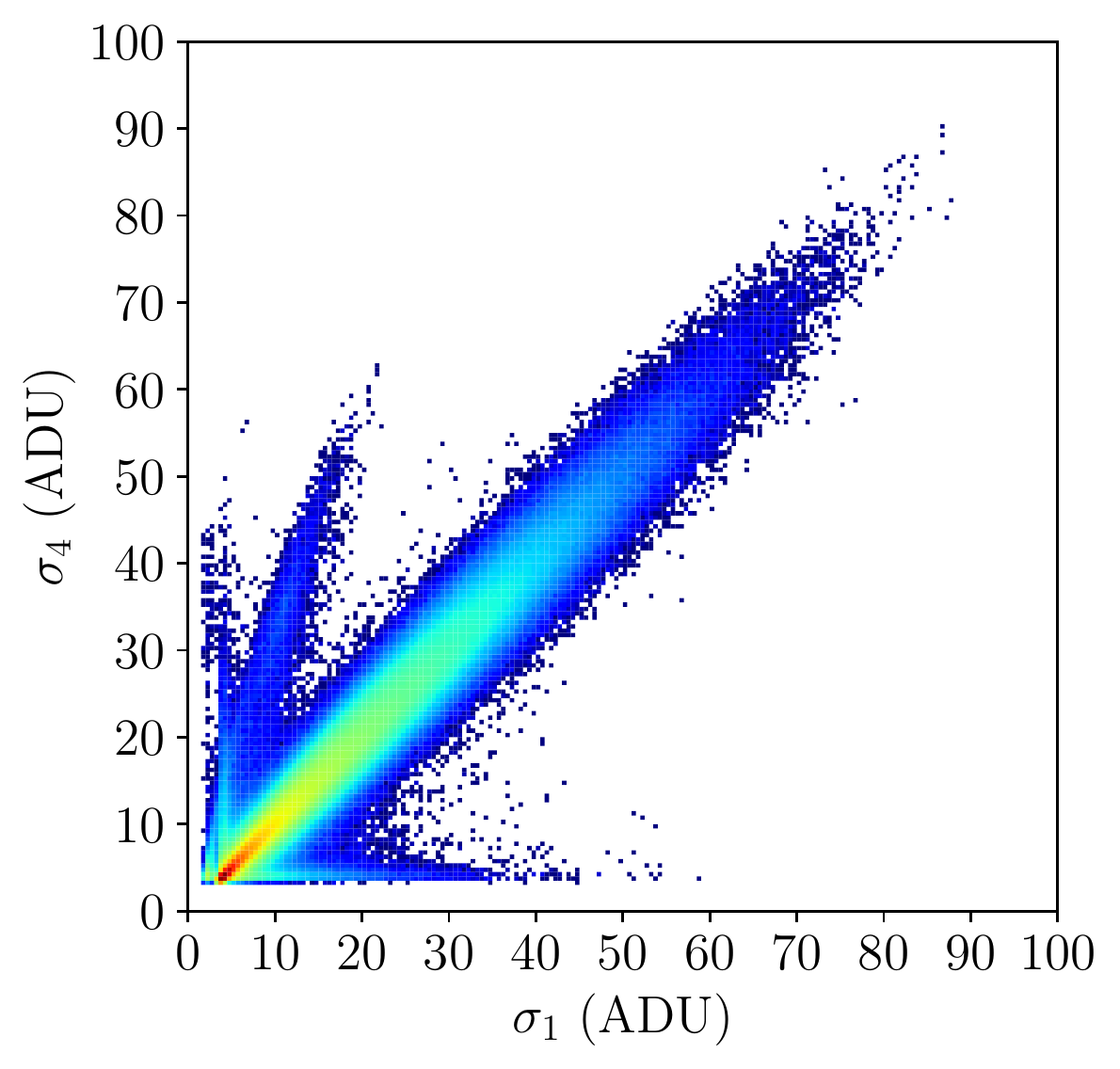}
}
\caption{Pixel-by-pixel comparisons of standard deviations of ``master bias''
images 1 (obtained on on May 27, 2022), 2 (obtained on May 26, 2022), 3
(obtained on May 20, 2022) and 4 (obtained on November 22, 2021), all obtained
by the same detector of telescope 4.  (All comparisons are with respect to
master bias image 1.)  Displays are logarithmic heat maps, with red
representing largest values and blue representing smallest values.}
\end{figure}

The distribution of the pixel-by-pixel standard deviations of one of the master
bias images from Figure 5 is shown in Figure 7, expressed in electron units
adopting the nominal gain 0.27 e$^-$ ADU$^{-1}$ described in \S\ 4.1.1 and
displayed on linear and logarithmic scales.  From Figure 7 it is apparent that
(1) the lowest pixel noise levels are $\approx 0.8$ e$^-$ and (2) the
distribution of the pixel noise levels has a long tail that extends to nearly
25 e$^-$.  The mode and median of the pixel noise level distribution pixel are
each 1.2 e$^-$, which is significantly lower than nominal value 1.5 e$^-$ of
the read noise described in \S\ 4.1.1.  Similar results are obtained for
different master bias images and for master bias images of different detectors.

\begin{figure}[ht!]
\centering
\subfloat{
  \includegraphics[width=0.45\linewidth]{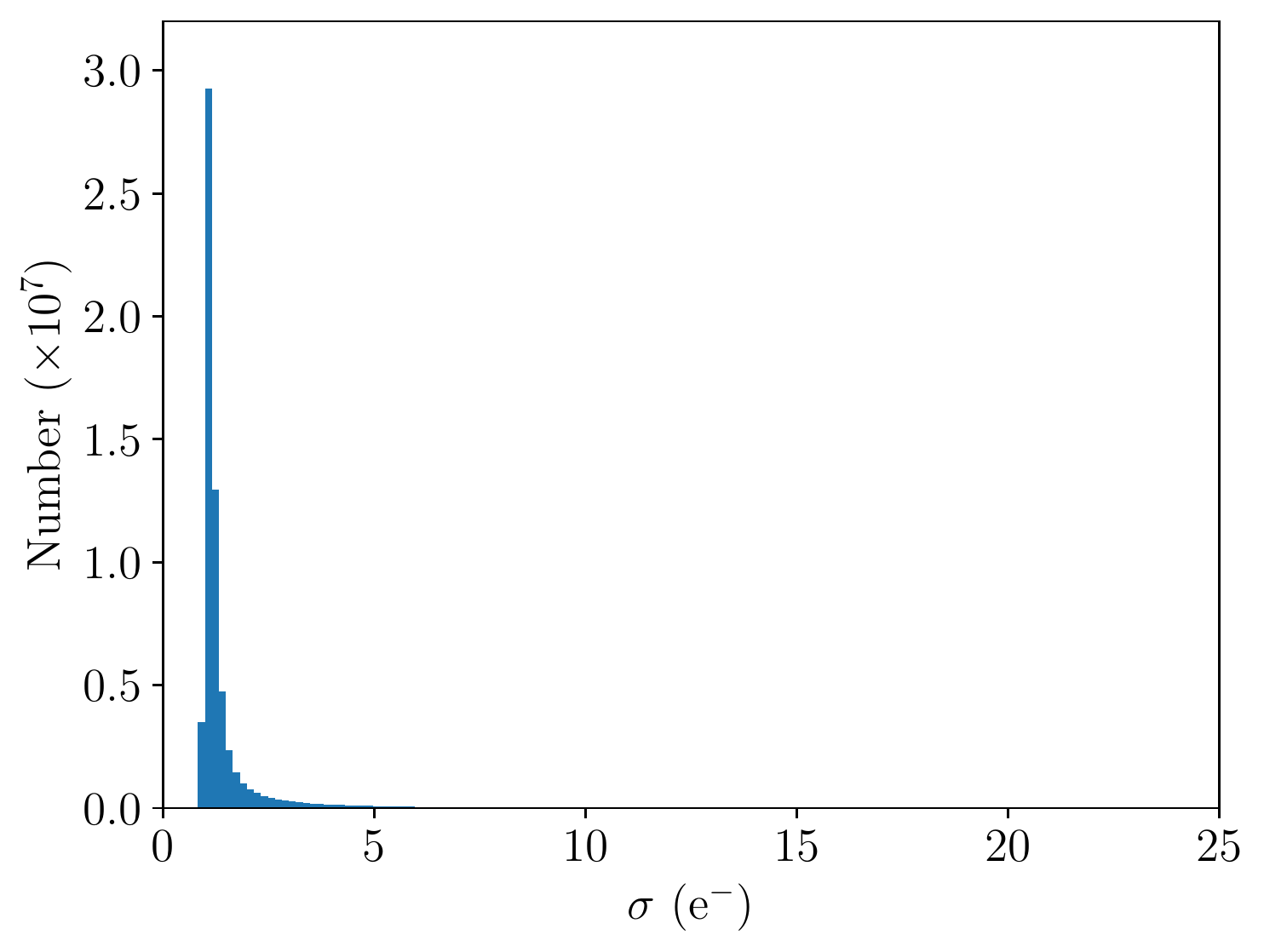}
}
\subfloat{
  \includegraphics[width=0.45\linewidth]{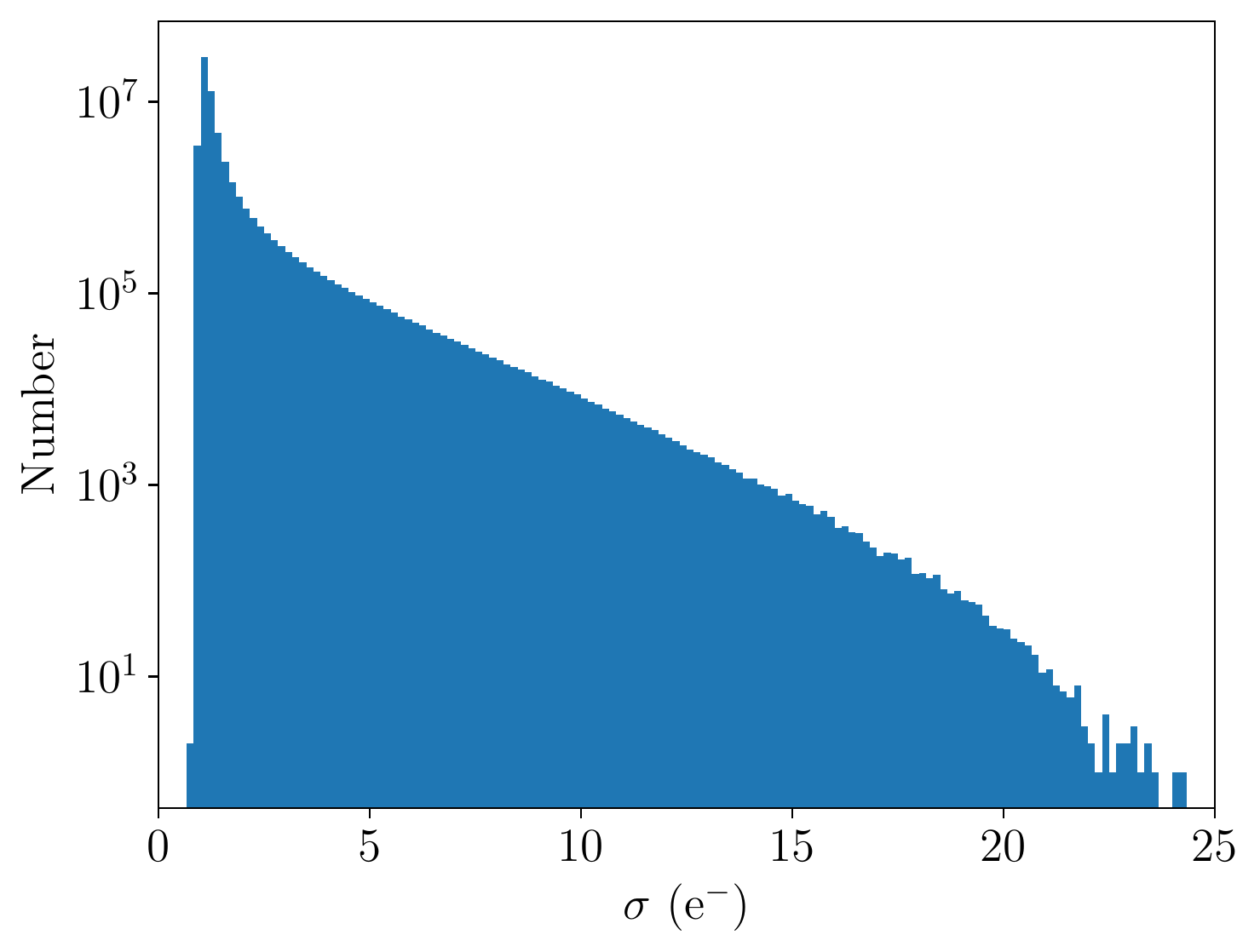}
}
\caption{Distribution of pixel-by-pixel standard deviations of master bias
image 1 from Figure 3, expressed in electron units and shown on ({\em a, left})
a linear scale and ({\em b, right}) a logarithmic scale.  Mode and median of
the distribution are each 1.2 e$^-$.  Bin widths are 0.2 e$^-$.}
\end{figure}

It is well known that CMOS detectors exhibit ``random telegraph noise'' (RTN),
which is evident as a long tail in the distribution of pixel noise levels of
bias or dark images that extends to many times (up to 10 or 100) the median of
the distribution \citep[e.g.][]{cha2019}.  This RTN is a phenomenon that
results from capture and release of electrons by defects in the gate oxide of
the detector.  We attribute the long tail of the distribution of Figure 7b to
RTN, or in other words, we postulate that the pixels that exhibit a pixel noise
level significantly in excess of the median 1.2 e$^-$ of the distribution
result from RTN.  Roughly 12\% of the pixels of the master bias image from
which Figure 7 was constructed exhibit a pixel noise level in excess of 2
e$^-$, which drops to 6\% in excess of 3 e$^-$, 2\% in excess of 5 e$^-$, and
0.3\% in excess of 10 e$^-$.

But we see no reason to make a conceptual distinction between pixels that
exhibit ``read noise'' and pixels that exhibit ``random telegraph noise.''  The
pixel noise level distribution of Figure 5 is unimodal rather than bimodal, and
we believe that any such distinction would be arbitrary.  Accordingly, we say
that the ``read noise'' of the detector is the value 1.2 e$^-$ of the median
(and mode) of the distribution but that some small fraction of pixels exhibit
pixel noise levels significantly in excess of the read noise.  Because we
characterize the pixel noise level of each pixel individually and because the
pixel noise level is very stable (as discussed above), the RTN has little
effect in practice.  Pixels that are significantly affected by RTN can be
included into any processing weighted by their associated pixel noise levels
with no detrimental effect, or such pixels can be identified and excluded from
processing (say for purposes of visual display).

\subsubsection{Gain}

We determined the gain of each CMOS detector by measuring pixel-to-pixel
fluctuations in flattened quotients of pairs of images of the twilight sky.
(See \S\ 4.5 for a brief discussion of our field flattening and background
subtraction techniques.)  We found a gain of 0.26 e$^-$ ADU$^{-1}$ for each
CMOS detector, which is very close to the nominal gain 0.27 e$^-$ ADU$^{-1}$
described in \S\ 4.1.1.

\subsubsection{Linearity}

We assessed the linearity of each CMOS detector by measuring the gain of the
detector (as described in \S\ 4.1.3) as a function of the ``illumination'' (or
the average signal level across the detector) in flattened quotients of pairs
of images of the twilight sky with similar values of illumination.  We found
that the gain increases by $\approx 0.38\%$ as the illumination varies from
54,000 to 57,000 ADU.  An example of gain versus illumination is shown in
Figure 8.  We defer a more detailed assessment of linearity until later.

\begin{figure}[ht!]
\centering
\includegraphics[width=0.45\linewidth]{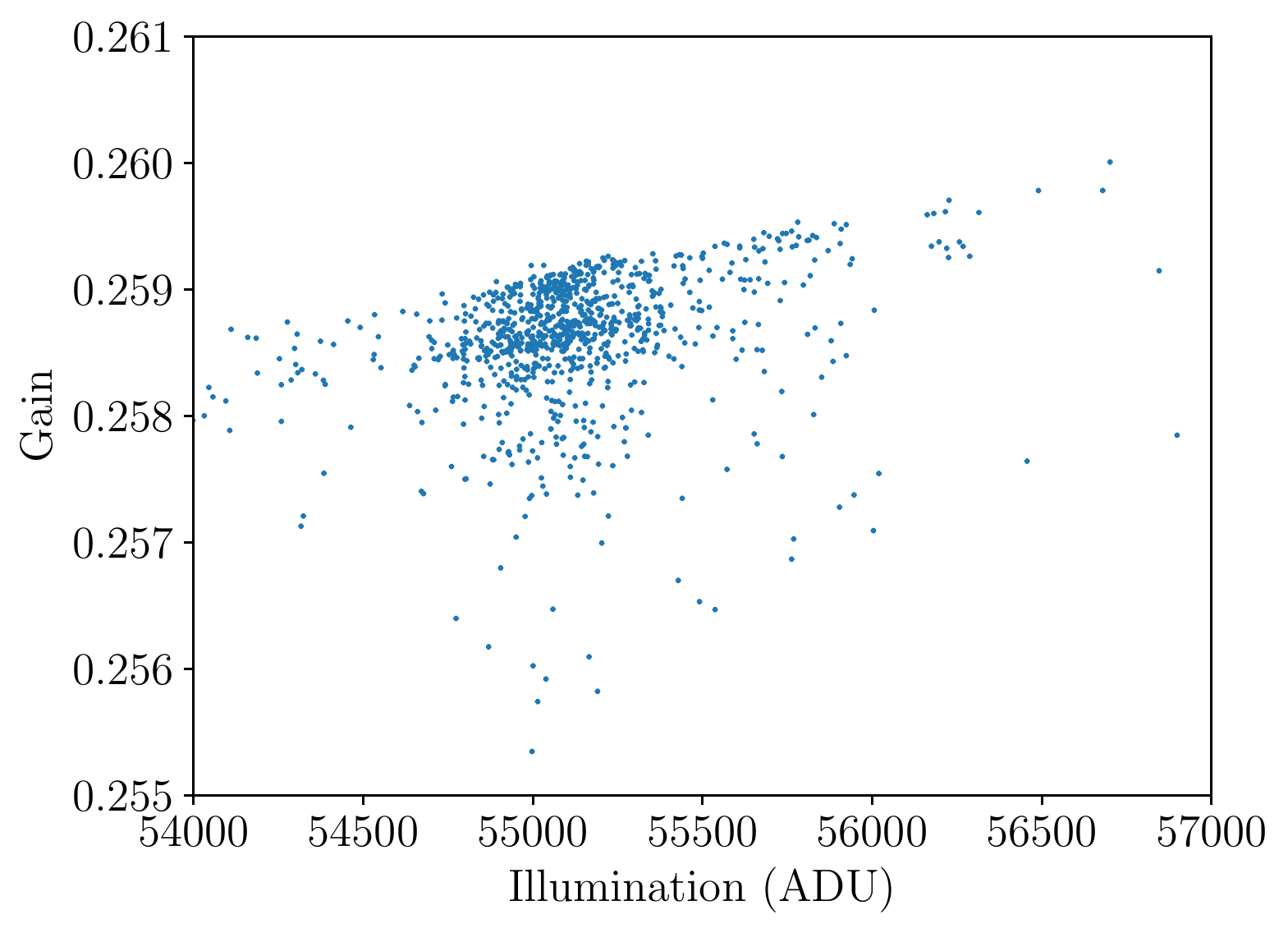}
\caption{Gain versus illumination measured from flattened quotients of pairs of
images of twilight sky obtained by telescope 4 in spring 2022.  Here
``illumination'' is average signal level across detector.  Measured points
trace out an upper envelope that indicates that gain increases by $\approx
0.38\%$ as illumination varies from 54,000 to 57,000 ADU.  Points that deviate
below predominant trend of gain versus illumination arise due to contributions
to the pixel-to-pixel fluctuations of twilight-sky quotients in excess of
photon statistics (like clouds).  We take upper envelope as indicative of
relationship between gain and illumination.}
\end{figure}

\subsubsection{Dark Current}

We determined the dark current of each CMOS detector by the measuring the mean
values of bias-subtracted dark images of exposure 3600 s.  We found a dark
current of typically $1.4 \times 10^{-3}$ e$^-$ s$^{-1}$ pix$^{-1}$ for each
CMOS detector at a detector temperature of $\approx -12$ C.  There are no
obvious significant patterns or structures in the dark images.

\subsubsection{Image Persistence}

We searched for evidence of ``image persistence''---or the tendency of some
CMOS detectors to capture and retain charge when illumination levels are high
and to then gradually release this charge over time \citep{kar2021}---by
comparing back-to-back dithered images of bright stars.  The same portions of
two 100 s dithered images of Vega obtained 11 s apart are shown in Figure 9.
There is no evidence of persistence of the image of Vega from the first
exposure to the second exposure.  Similar results are obtained for other images
of bright stars and for images of bright stars obtained by the other
telescopes.  We conclude that images obtained by the CMOS detectors are not
significantly affected by image persistence.

\begin{figure}[ht!]
\centering
\subfloat{
  \includegraphics[width=0.25\linewidth, angle=-90, origin=c]{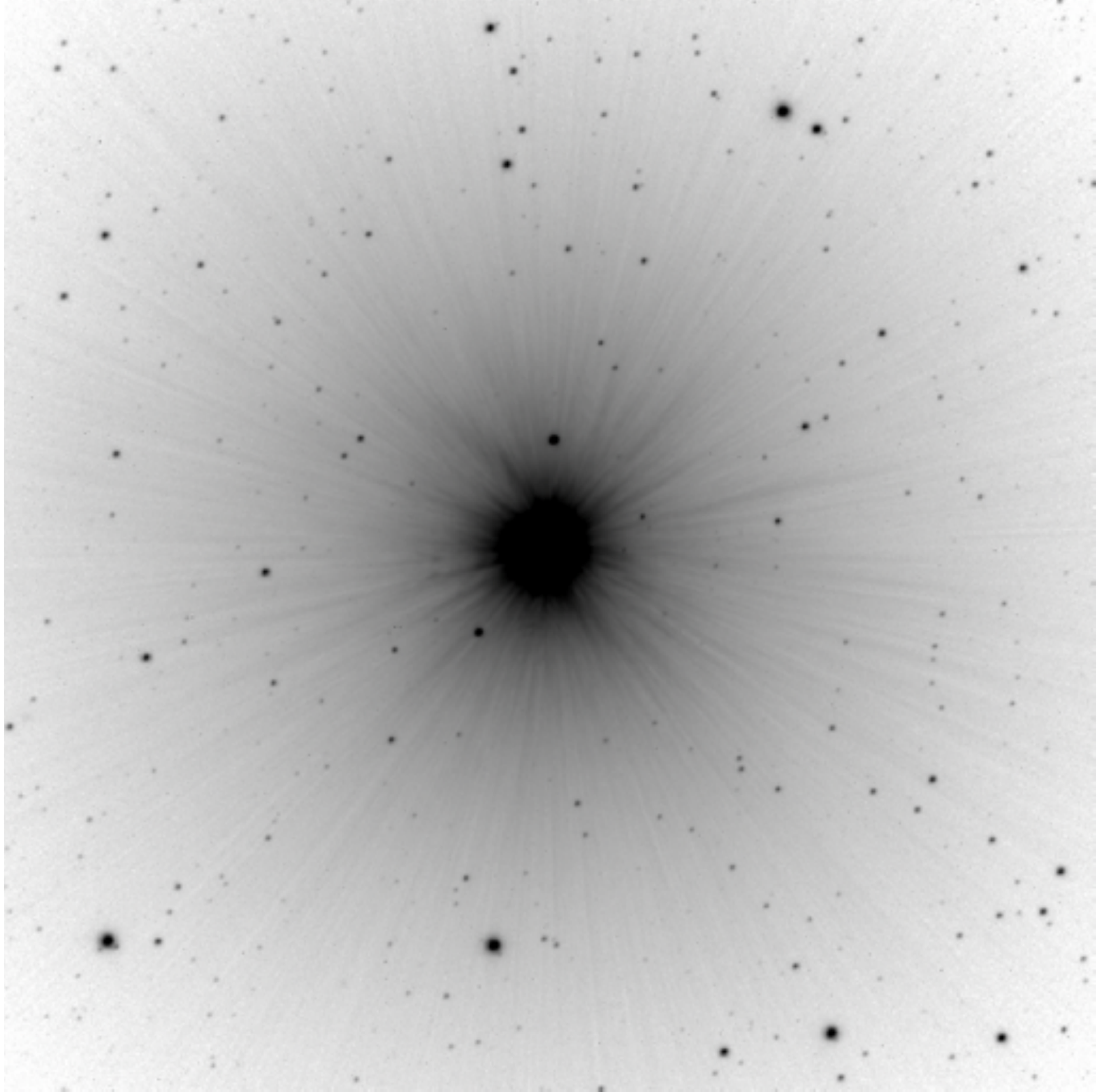}
}
\subfloat{
  \includegraphics[width=0.25\linewidth, angle=-90, origin=c]{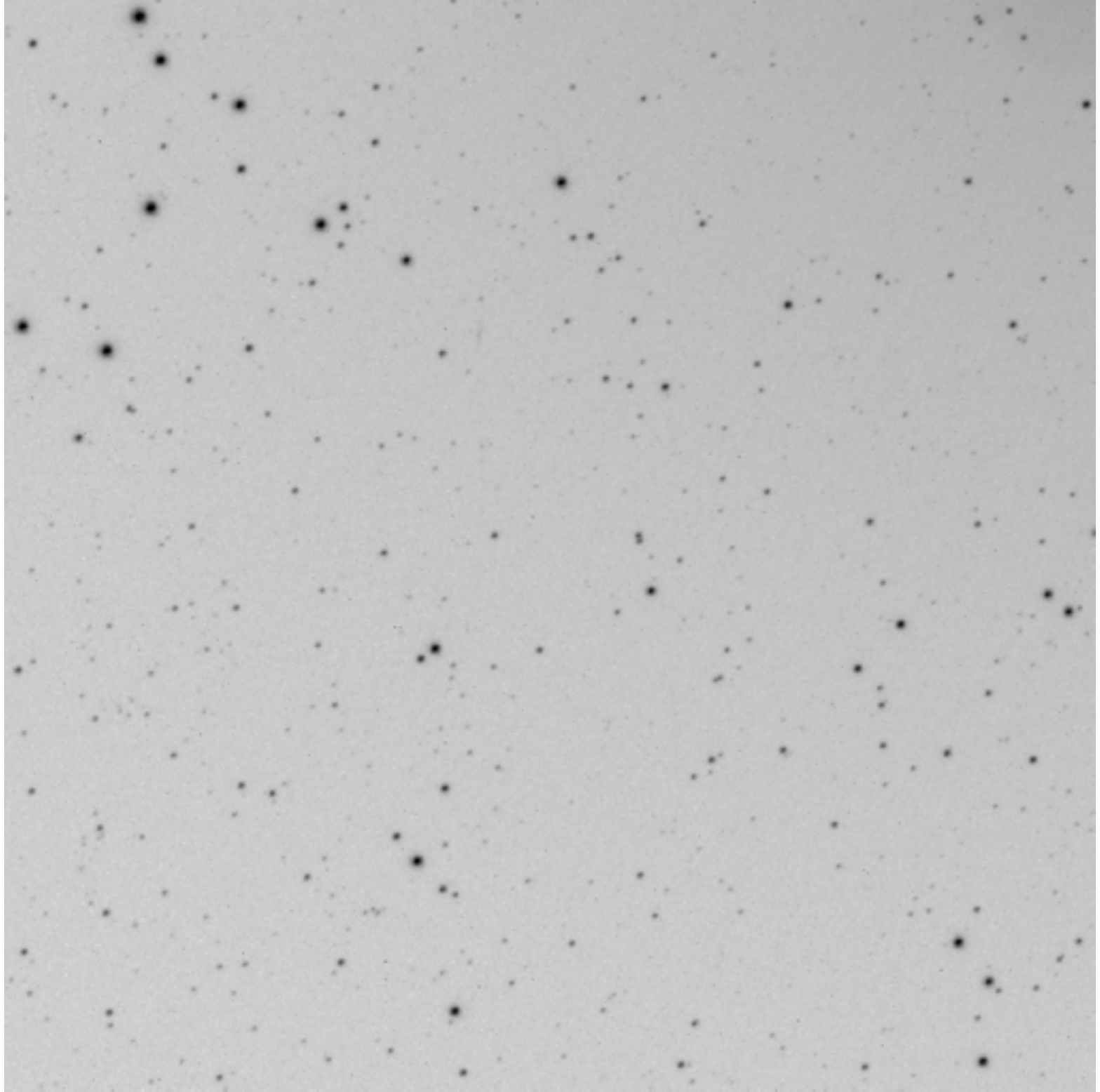}
}
\subfloat{
  \includegraphics[width=0.25\linewidth, angle=-90, origin=c]{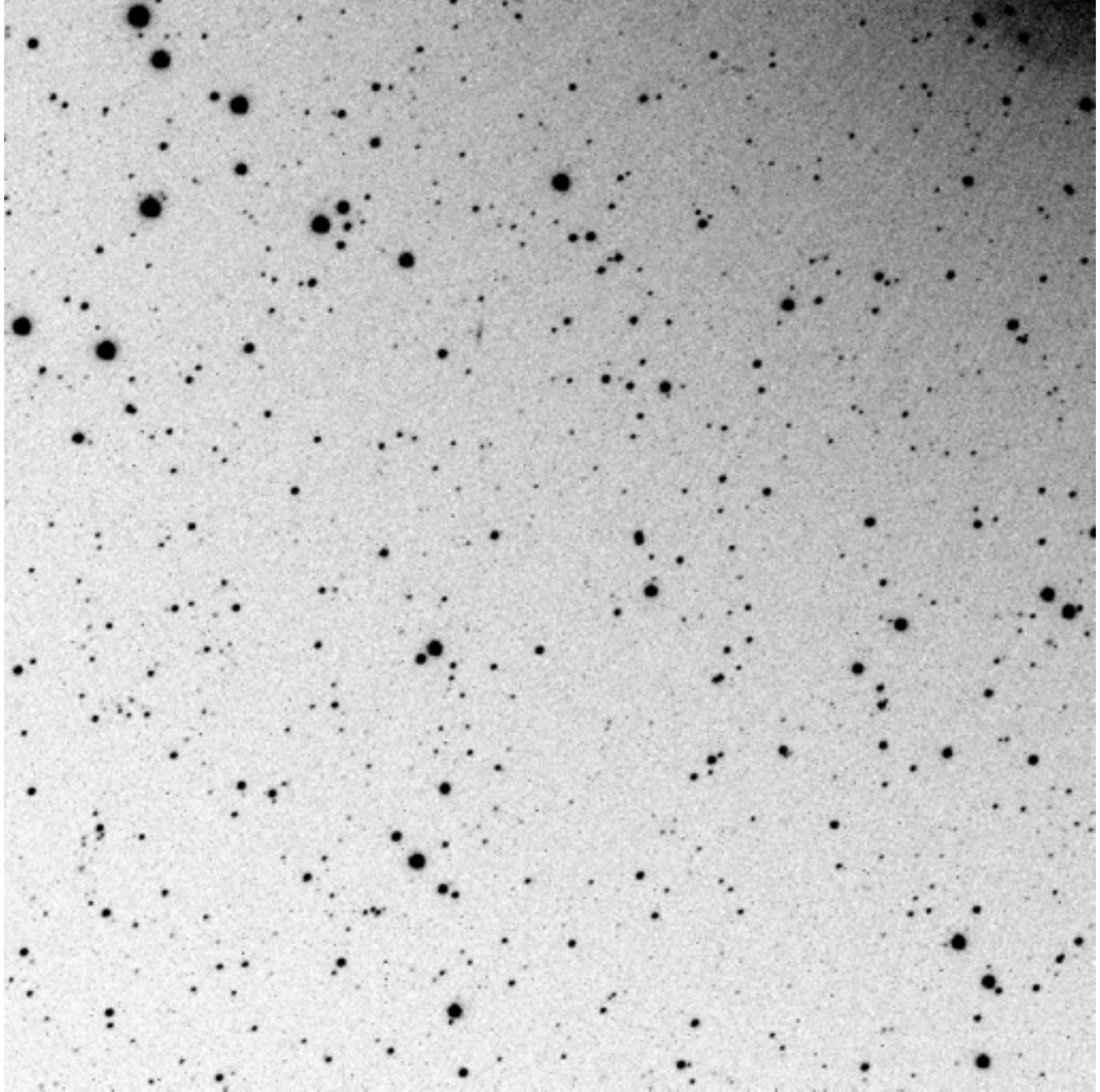}
}
\caption{Images of Vega and vicinity obtained by telescope 2 on June 7, 2022:
({\em a, left}) image obtained at 08:23:19 UTC with exposure of 100 s on
logarithmic stretch, ({\em b, center}) image obtained at 08:25:10 UTC with
exposure of 100 s on logarithmic stretch, ({\em c, right}) same image as in
{\em b}, but on linear stretch.  In all cases, a $1000 \times 1000$ pix$^2$
portion of the image is shown, centered on the location of Vega in the first
image.  There is no evidence of persistence of the image of Vega from the first
exposure to the second exposure.}
\end{figure}

\subsubsection{Cosmic Ray Events}

We assessed the incidence of cosmic ray events by processing images through
MaxiMask \citep{pai2020}, which is a convolutional neural network that detects
contaminants in astronomical images.  Applied to a 60-s exposure, this
processing identifies typically 20,000 events that are classified as cosmic ray
events, each extending over typically 12 pixels and in total affecting $\approx
0.4\%$ of the pixels comprising the image.  But these events involve very
little charge (typically $\lesssim 100$ e$^{-}$ each), and no pixels are ever
close to saturated by these events, and these events are always very localized.
The incidence of these events is much higher than is expected for cosmic ray
events, and we are uncertain as to the nature of these events and whether few,
some, many, or most of these events are indeed produced by cosmic rays.  But
this processing nevertheless establishes an upper limit to the incidence of
cosmic ray events.  We conclude that images obtained by the CMOS detectors are
not significantly affected by cosmic ray events.

\subsubsection{Cosmetic Defects}

We found the CMOS detectors to be notably unaffected by the various
``cosmetic'' defects typical of CCD detectors, including hot and dark pixels
and charge bleeding from saturated stars.

\subsubsection{Summary}

Several of us have extensive experience working with CCD detectors dating back
many decades.  Our overall impression is that the Sony IMX455 back-illuminated
detector performs superbly.  We are particularly impressed with the very large
format, the very low read noise, the low dark current, the 16-bit ADC, the high
quantum efficiency, the very rapid read time, the long-term stability, and the
absence of cosmetic defects.  We believe that CMOS detectors should and will
soon replace CCD detectors in most or all astronomical applications.

\newpage

\subsection{Plate Scale and Field of View}

Considering the 3.76 $\mu$m pixel size of the $9576 \times 6388$ format Sony
IMX455 detector together with the nominal effective focal length 907 mm of the
TEC 180 mm telescope with the A-P 0.72x four-element telecompressor corrector
yields a nominal plate scale of 0.86 arcsec pixel$^{-1}$ and a nominal field of
view $2.29 \times 1.53$ deg$^2$.  The measured values after astrometric
calibration are very close to the nominal values.  (See \S\ 4.4 for a brief
discussion of our astrometric calibration techniques.)  The solid angle
subtended by the field of view is 3.50 deg$^2$.  Details of the plate scale and
field of view are summarized in Table 3.

\begin{table}[ht]
\centering
\hspace{-0.35in}
\begin{tabular}{p{2.5in}c}
\multicolumn{2}{c}{{\bf Table 3:}  Details of Plate Scale and Field of View} \\
\hline
\hline
Plate scale \dotfill & 0.86 arcsec pix$^{-1}$ \\
Field of view \dotfill & $2.29 \times 1.53$ deg$^2$ \\
Solid angle subtended by field of view \dotfill & 3.50 deg$^2$ \\
\hline
\end{tabular}
\end{table}

The Rayleigh limit of a 180 mm-diameter aperture ranges from $\approx 0.5$
arcsec at wavelength $\lambda = 350$ nm to $\approx 1.0$ arcsec at $\lambda
= 700$ nm, which when convolved with a seeing profile of width, say, ${\rm
FWHM} = 1.2$ arcsec typical of a good astronomical site yields a point-spread
function (PSF) of width ranging from ${\rm FHWM} \approx1.3$ arcsec at $\lambda
= 350$ nm through ${\rm FWHM} \approx 1.6$ arcsec at $\lambda = 700$ nm.  The
0.86 arcsec plate scale of the telescope Nyquist samples a PSF of ${\rm FWHM}
\approx 1.7$ arcsec, hence Condor slightly undersamples the PSF under seeing
conditions typical of a good astronomical site, by a factor ranging from as
much $\approx 25$\% at $\lambda = 350$ nm through as little as $\approx 6$\% at
$\lambda = 700$ nm.  We conclude that Condor can obtain nearly Nyquist-sampled
images at a good astronomical site.

\subsection{Detector Uniformity and Vignetting}

An image of the twilight sky obtained with one of the six telescopes is shown
in Figure 10.  From Figure 10 it is apparent that:  (1) The detector response
is very uniform over the field of view, and the detector exhibits no obvious
blemishes or cosmetic flaws.  (As is expected, several out-of-focus dust motes
are clearly visible in the image of Figure 10.)  (2) The image circle is well
centered on the detector, which indicates that the optical beam is well
collimated.  And (3) the detector is nearly but not quite fully illuminated by
the image circle, with vignetting evident at the corners of the image.  Similar
results are obtained for other twilight images and for twilight images obtained
by the other telescopes.

\begin{figure}[ht!]
\centering
\includegraphics[width=0.40\linewidth, angle=-90, origin=c] {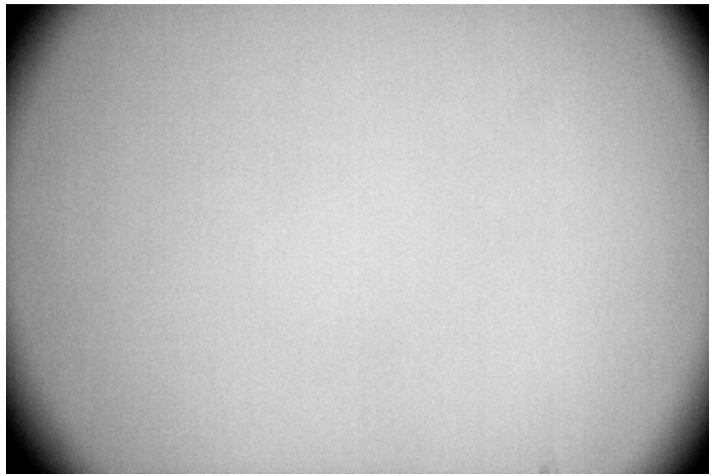}
\vspace{-0.5in}
\caption{Image of twilight sky obtained by telescope 4 near dawn on May 30,
2022.  Several out-of-focus dust motes are clearly visible in the image.}
\end{figure}

A diagonal cut across the image of Figure 10 running from the upper left corner
to the lower right corner is shown in Figure 11.  From Figure 11 it is apparent
that the illumination of the detector is very flat across most of the field of
view, with vignetting again evident at the corners of the image.  The maximum
vignetting at the very corners of the image reaches $\approx 15$\%.  Similar
results are obtained for other twilight images and for twilight images obtained
by the other telescopes.

\begin{figure}[ht!]
\centering
\includegraphics[width=0.45\linewidth]{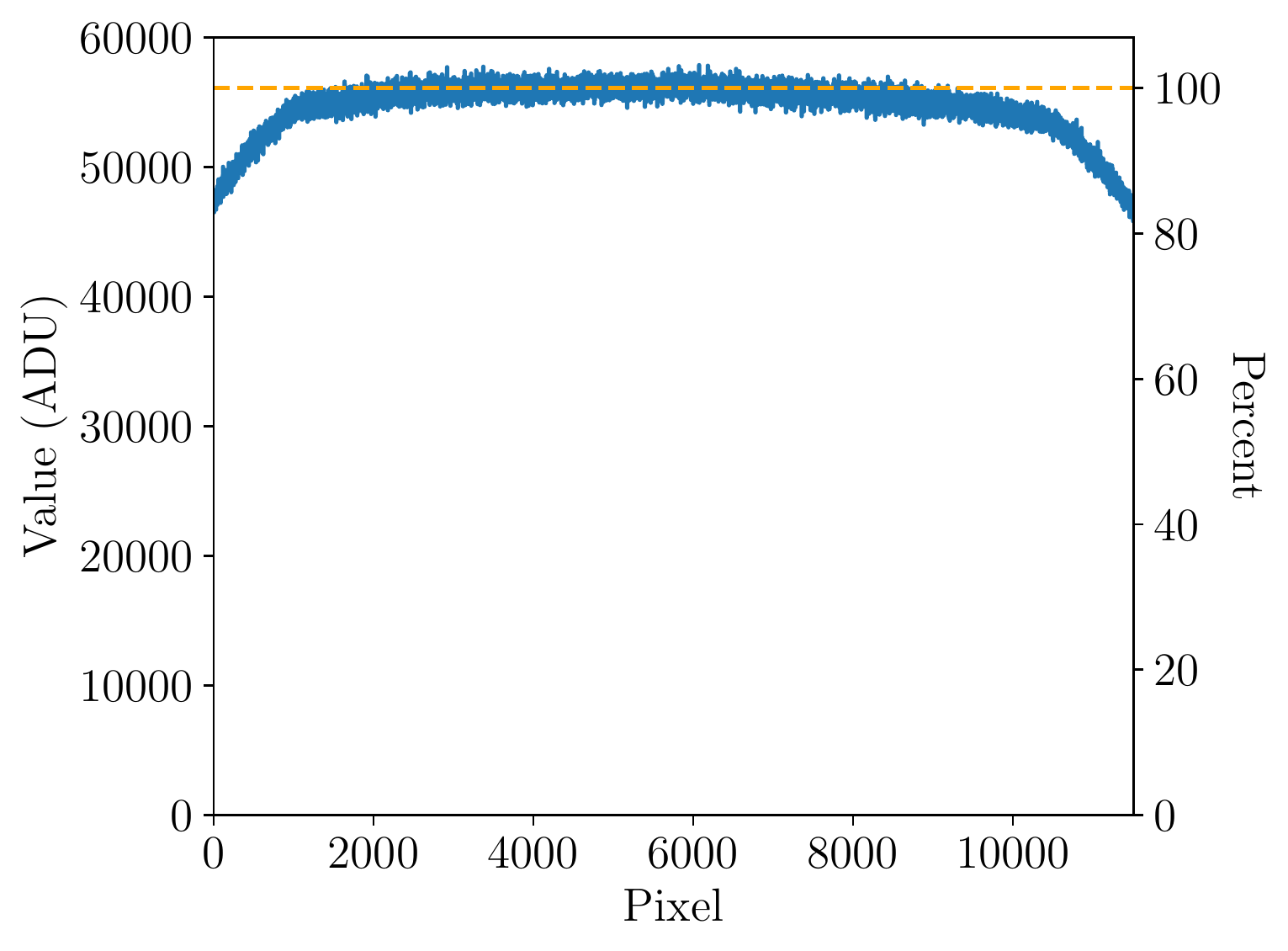}
\caption{Diagonal cut across image of Figure 9 running from upper left corner
to lower right corner.  Dashed orange line is set to peak mean value near
center of image.}
\end{figure}

\subsection{Astrometric Calibration}

Details of the astrometric calibration of Condor images will be described
elsewhere, but here we present a brief summary of our procedures and results.

Accurate astrometric calibration is crucial for all of the science objectives
of Condor.  Yet determining an accurate astrometric calibration over the wide
field of view of Condor is not straightforward, especially in the presence of
geometric distortions and field curvature at the corners of the images.
Accordingly, we developed a technique of measuring astrometric calibration that
is appropriate to Condor images.  Briefly, the technique proceeds as follows:
First, we start with a science image for which bias subtraction and field
flattening and background subtraction have already been performed.  Next, we
process the image through SExtractor \citep{ber1996}, which is a program that
identifies sources in astronomical images.  This processing produces a list of
pixel coordinates of sources detected in the image.  Next, we query the Gaia
DR3 catalog \citep{gai2017, gai2018, gai2021} to obtain a list of sources in
the field, noting in particular the celestial coordinates and proper motions of
the sources.  Next, we fit for parameters of the affine transformation that
best matches the pixel and celestial coordinates (corrected for proper
motions) of a subset of the sources, and we construct a list of matched pixel
and celestial coordinates of the sources.  Finally, we fit for parameters of a
higher-order model that best matches the pixel and celestial coordinates,
taking as parameters the affine transformation parameters and the coefficients
of a seventh-order geometric distortion polynomial including radial terms, in
the ``TPV'' projection.  (The TPV projection is similar to the TAN projection
described by \citealp{cal2002} but includes a seventh-order geometric
distortion polynomial.)

By applying this technique, we consistently obtain an astrometric calibration
with systematic differences between the transformed pixel and celestial
coordinates of $\lesssim 0.1$ arcsec.

\subsection{Field Flattening and Background Subtraction}

Details of the field flattening and background subtraction of Condor images
will be described elsewhere, but here we present a brief summary of our
procedures and results.

Field flattening and background subtraction are critical to many of the science
objectives of Condor.  Yet there are no obvious flat sources of illumination to
which Condor could be pointed to produce field-flattening calibration images.
We experimented with producing ``sky flat'' images by median combining
unregistered science images (after scaling to account for difference in
background level).  But we note two very significant problems with this
approach:  (1) the night sky is itself not flat,  and (2) photon noise in a sky
flat image can never be made insignificant in comparison to photon noise in the
(combined) science images, because the number of photons recorded by the sky
flat image grows only in direct proportion to the number of photons recorded by
the science images.

Instead, we have developed a technique of field flattening and background
subtraction that is based on forming a quotient with a twilight image.
Briefly, the technique proceeds as follows:  First, the science image is
divided by a twilight image (obtained by the same telescope with the same
detector).  This quotient is devoid of all instrumental effects (including all
telescope and detector effects) but contains the signatures of the (different
and unknown) sky backgrounds of the science and twilight images.  Next, the
quotient is fitted by a high-order (typically eighth-order) two-dimensional
polynomial, rejecting regions around sources present in the science image.
This polynomial fit accounts for the backgrounds of both the science and
twilight images.  Finally, the polynomial fit is subtracted from the quotient.
This sequence of steps produces a version of the science image that is
simultaneously corrected for field flattening and background subtraction.

By experimenting with pairs of twilight images (i.e.\ by substituting another
twilight image for the science image), we found that we are able to carry out
field flattening and background subtraction to within fundamental physical
limitations.  In particular, we found that the residuals of the quotients after
the polynomials are subtracted result from (1) rapidly-varying (i.e.\ with
times scales less than $\approx 10$ s) stochastic fluctuations of the
atmosphere on scales of several hundred pixels and (2) photon statistics on
scales of one pixel.  We used this technique, for example, to determine the
gain of each CMOS detector by measuring photon statistics on scales of one
pixel of pairs of twilight images, as described in \S\ 4.1.3.

\subsection{Photometric Calibration}

Details of the photometric calibration of Condor images will be presented
elsewhere, but here we present a brief summary of our procedures and results.

Exquisite photometric calibration is not critical to many of the science
objectives of Condor.  Accordingly, our current efforts have concentrated on
obtaining an approximate photometric calibration suitable for quantifying
sensitivity and for identifying variable and transient sources.  Briefly, the
technique proceeds as follows:  First, we start with science images for which
bias subtraction, astrometric calibration, and field flattening and background
subtraction have been performed.  Next, we query the Gaia DR3
catalog \citep{gai2017, gai2018, gai2021} to obtain a list of sources in the
field, noting in particular the celestial coordinates and $g'$ magnitudes of
the sources.  Next, we use aperture photometry techniques to measure
brightnesses of sources in the list of sources, excluding very bright or very
faint sources.  Finally, we compare the measured brightnesses to the $g'$
magnitudes across the ensemble to derive a rough photometric calibration.

The distributions of the ``photometric calibration'' (i.e.\ the energy flux
density that corresponds to 1 ADU s$^{-1}$) and the ``magnitude zero point''
(i.e. the $g'$ magnitude that corresponds to 1 ADU s$^{-1}$) measured for a
large number of images obtained through the luminance filter are shown in
Figure 12.  From Figure 12 it is apparent that under the clearest conditions
ever experienced by Condor, the photometric calibration is $\approx 3.5$
$\mu$Jy ADU$^{-1}$ s and the magnitude zero point $m_0$ is $m_0 \approx 22.7$
through the luminance filter.  Measurements of the photometric calibration and
magnitude zero point through other filters are given in Appendix A.

\begin{figure}[ht!]
\centering
\subfloat{
  \includegraphics[width=0.45\linewidth]{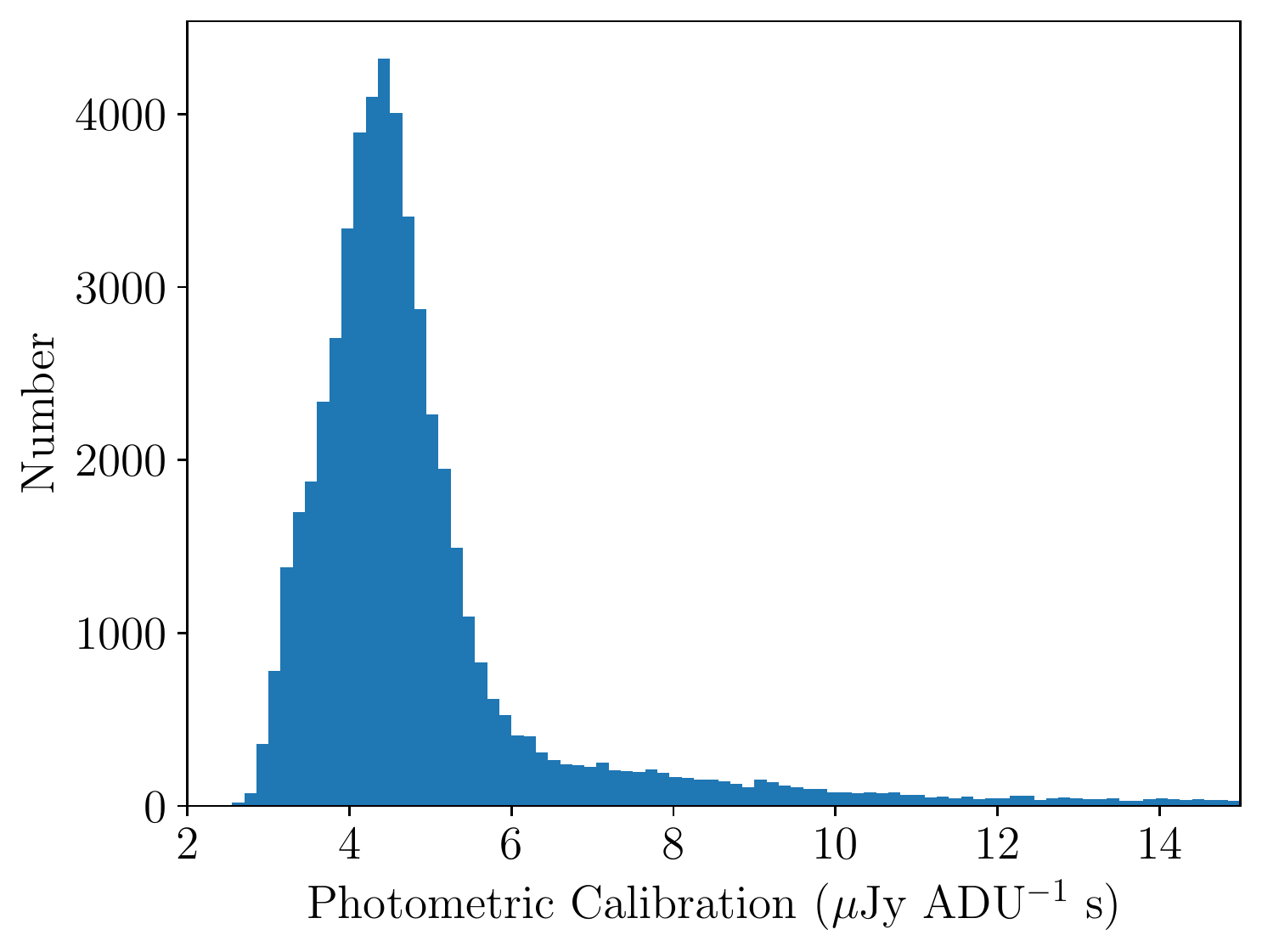}
}
\subfloat{
  \includegraphics[width=0.45\linewidth]{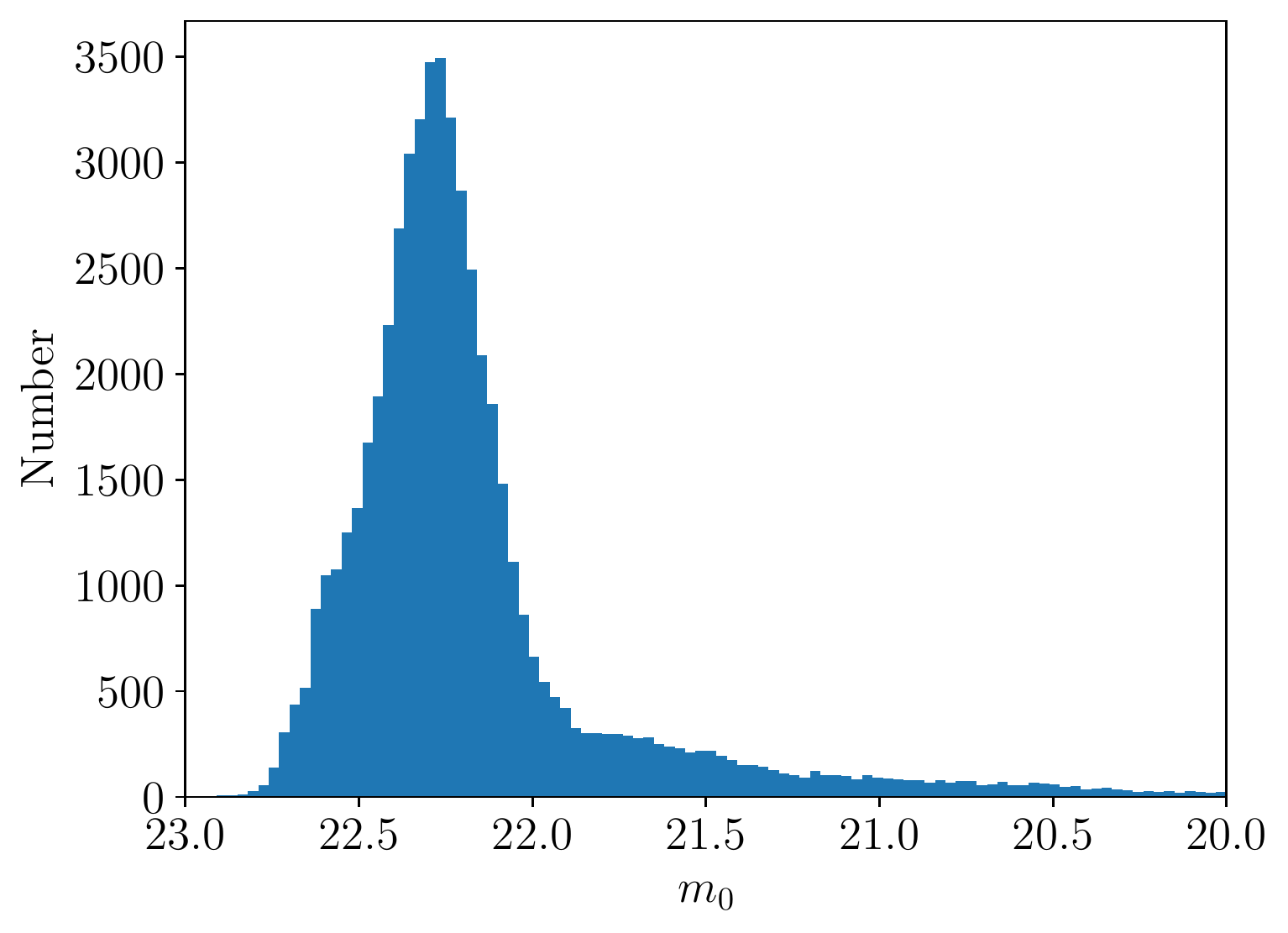}
}
\caption{Distributions of ({\em a, left}) ``photometric calibration'' (i.e.\
energy flux density that corresponds to 1 ADU s$^{-1}$) and ``magnitude zero
point'' $m_0$ (i.e.\ $g'$ magnitude that corresponds to 1 ADU s$^{-1}$) of
images obtained through luminance filter over time interval stretching from
April 21, 2021 through June 19, 2022.  Bin width of left panel is 1.5 $\mu$Jy
ADU$^{-1}$ s, and bin width of right panel is 0.03 mag.}
\end{figure}

\subsection{Image Quality and Point-Spread Function}

An image of Vega obtained by one telescope is shown in Figure 13.  This image
demonstrates that the Condor PSF is exceptionally clean and that the only
apparent features of the PSF are (1) a very faint reflection visible to the
lower left of the star in the left panel and (2) very low-level ``spikes'' that
emanate radially from the star (which are more easily seen in the right panel).
Following \citetalias{abr2014}, we tentatively attribute these spikes to
diffraction from striae in the glass elements and lattice imperfections in the
crystalline CaF$_2$ elements of the lenses.  The left panel of Figure 13 may be
directly compared to Figure 8 of \citetalias{abr2014}, which is a similar image
of Venus on a logarithmic stretch over a scale of 50 arcmin.  (But we note that
Venus is, of course, not a point source.)  The Condor image appears
substantially cleaner than the Dragonfly image, and the Condor image does not
show the multiple reflections, ``ghosts,'' and halos evident in the Dragonfly
image.  Similar results are obtained for other images of Vega (and other bright
sources) and for images of Vega (and other bright sources) obtained by the
other telescopes.

\begin{figure}[ht!]
\centering
\subfloat{
  \includegraphics[width=0.37\linewidth, angle=-90, origin=c]{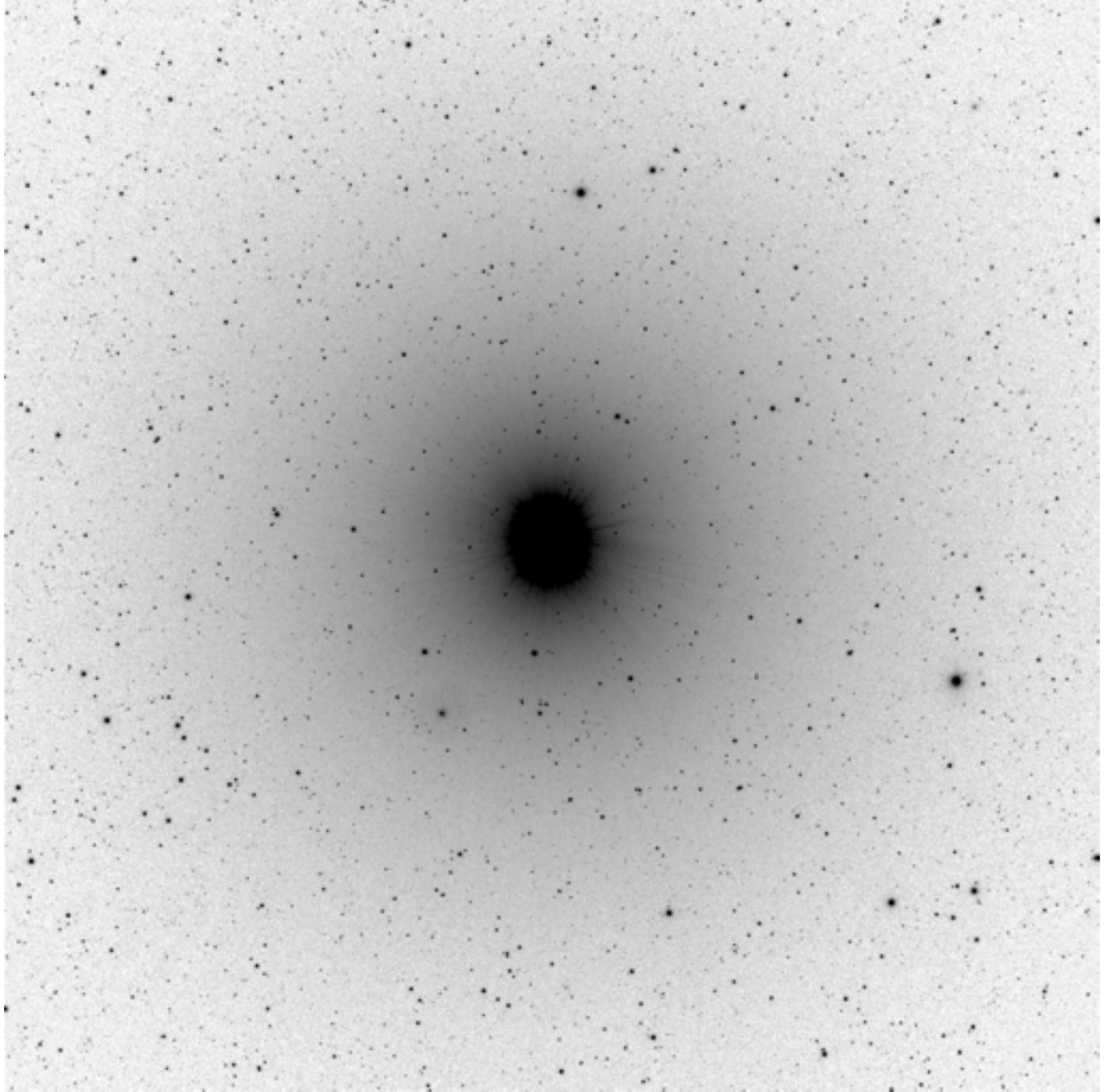}
}
\subfloat{
  \includegraphics[width=0.37\linewidth, angle=-90, origin=c]{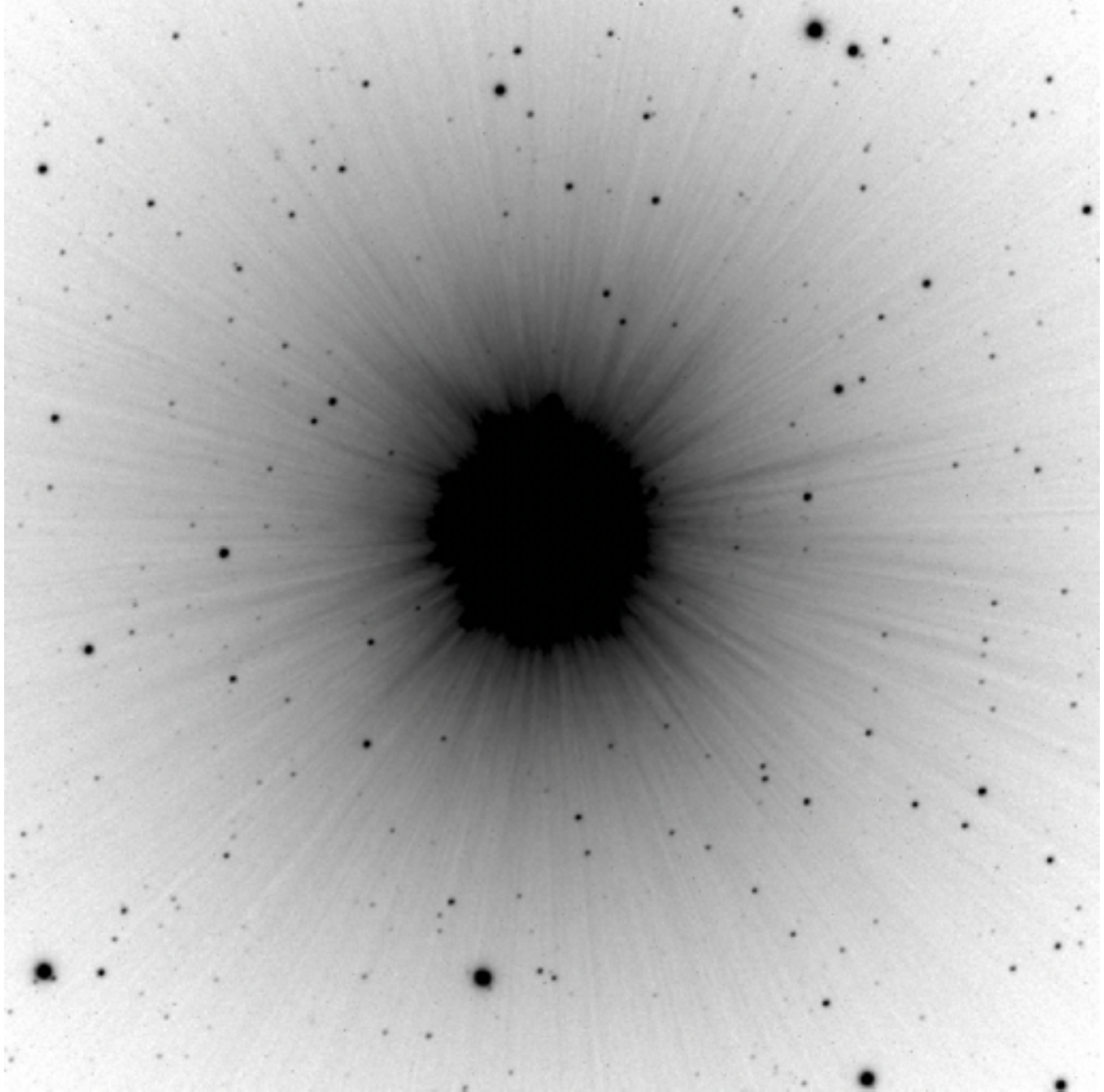}
}
\caption{Image of Vega obtained by telescope 2 on June 7, 2022 with an exposure
time of 100 s on a logarithmic stretch over a scale of ({\em a, left}) 50
arcmin on a side and ({\em b, right}) 12.5 arcmin on side.  This image
demonstrates that the Condor PSF is exceptionally clean and that the only
apparent features of the PSF are (1) a very faint reflection visible to the
lower left of the star in the left panel and (2) very low-level ``spikes'' that
emanate radially from the star (which are more easily seen in the right panel).
Following \citetalias{abr2014}, we tentatively attribute these spikes to
diffraction from striae in the glass elements and lattice imperfections in the
crystalline CaF$_2$ elements of the lenses.  The left panel may be directly
compared to Figure 8 of \citetalias{abr2014}, which shows a similar image of
Venus obtained by Dragonfly on a logarithmic stretch over a scale of
50 arcmin.}
\end{figure}

The Condor PSF pieced together from the image of Figure 13 and other similar
images of exposure times 0.1, 1, 10, and 100 s is shown in Figure 14.  In
Figure 14, the vertical axis gives surface brightness $\mu_{\rm PSF}$ of a
zero-magnitude source in units mag arcsec$^{-2}$, and the horizontal axis gives
angular distance $r$ in units arcsec on a logarithmic scale.  In Figure 14, the
PSF on scales $r > 10$ arcsec was estimated by determining the median values
within circular annuli centered on Vega.  The median values in ADU units were
converted to surface brightness units using the photometric calibration
described in \S\ 4.6, but otherwise these values were not scaled in any way.
This is appropriate because Vega is the primary photometric standard.  On
scales $r < 10$ arcsec the image of Vega was saturated even in the 0.1 s
exposure, so the PSF was estimated by determining the median values within
circular annuli of a fainter star.  Figure 14 also shows the Dragonfly PSF from
\citet{liu2022}.  The two PSFs are very similar on scales $r \lesssim $ 50
arcsec but appear to deviate on scales $r \gtrsim 50$ arcsec.  But it is
important to note that: (1) the Condor PSF on scales $r \lesssim 5$ arcsec is
set by seeing conditions and (2) both the Condor and Dragonfly PSFs on scales
$r \gtrsim 10$ arcsec depend on weather and atmospheric conditions.  We defer a
detailed comparison between the Condor and Dragonfly PSFs until later.

\begin{figure}[ht!]
\centering
\includegraphics[width=0.45\linewidth]{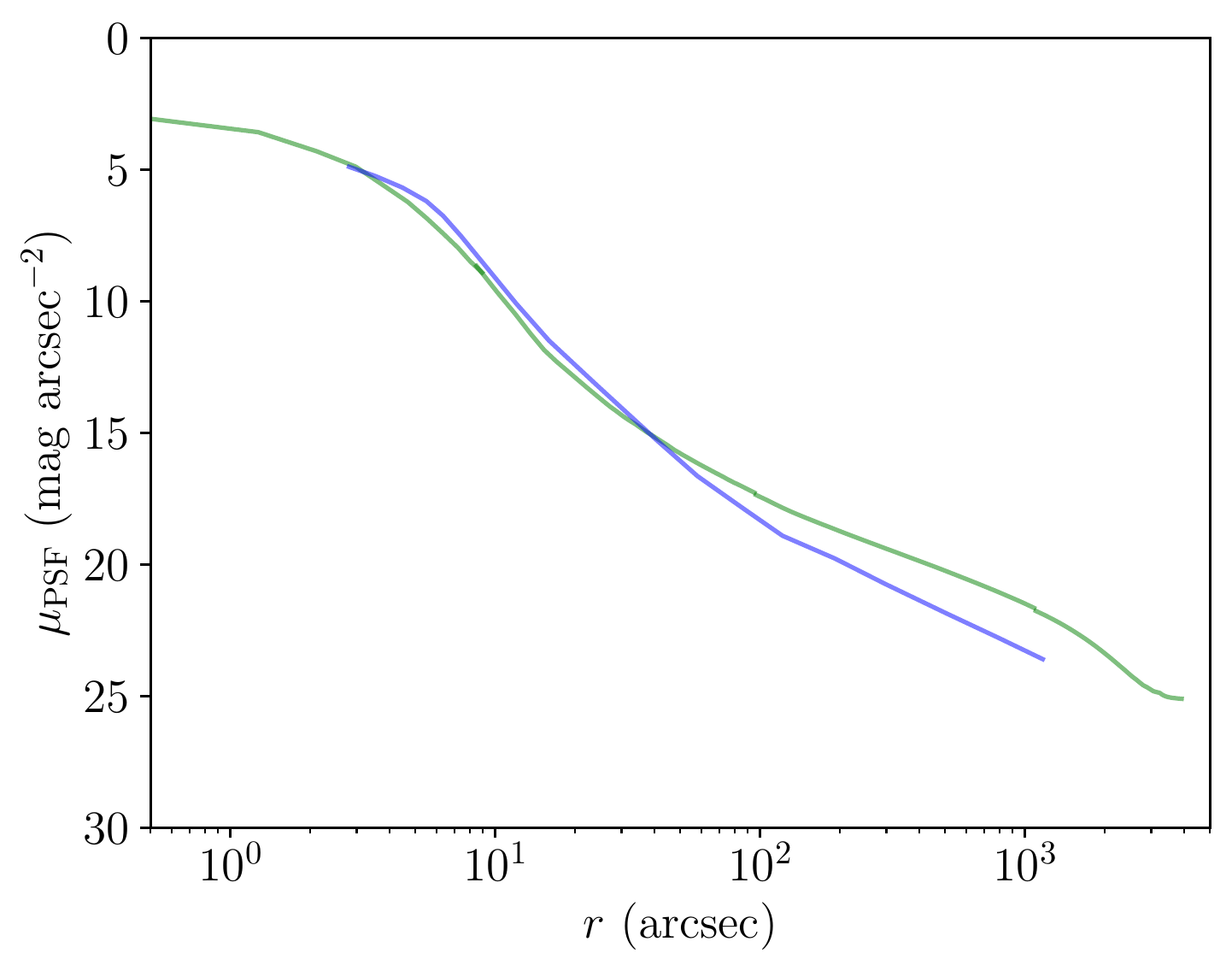}
\caption{Condor PSF (green) determined from image of Figure 13 and other similar
images of exposure times 0.1, 1, 10, and 100 s.  Vertical axis gives surface
brightness $\mu_{\rm PSF}$ of a zero-magnitude source in units mag
arcsec$^{-2}$, and horizontal axis gives angular distance $r$ in units arcsec
on a logarithmic scale.  Plot also shows Dragonfly PSF (blue) from
\citet{liu2022}.  The two PSFs are similar on scales $r \lesssim 50$ arcsec
but appear to deviate on scales $r \gtrsim 50$ arcsec.  But note that:  (1)
Condor PSF on scales $r \lesssim 5$ arcsec is set by seeing conditions and (2)
both Condor and Dragonfly PSFs on scales $r \gtrsim 10$ arcsec depend on
weather and atmospheric conditions.}
\end{figure}

The distribution of the ``image quality'' FWHM of a large number of images
obtained through the luminance filter are show in Figure 15.  We define the
``image quality'' FWHM of an image to be the FWHM of the central core of the
autocorrelation function of the image.  We found through experimentation that
the FWHM of the central core of the autocorrelation function of an image is
$\approx 2$ times the typical FWHM of point sources in the image.  Hence the
image quality FWHM is an easily-measured statistic that characterizes the
typical FWHM of point sources in the image.  We note that the typical FWHM of
point sources in the image is set by a combination of seeing, focus, tracking
errors, and field curvature (in the sense that focus may not be completely
uniform across the field).  Hence the distribution of Figure 15 represents the
sum of these effects.  In the months immediately following first light, when
methods of controlling focus, tracking errors, and field curvature were still
being developed, these effects contributed significantly to image quality.  But
by several months later, when these effects were better controlled, image
quality was set primarily by seeing.  The mode of the image quality FWHM
distribution is 1.6 arcsec, and the median of the distribution is 2.0 arcsec.
Because Condor Nyquist samples a PSF of ${\rm FWHM} \approx 1.7$ arcsec (as
described in \S\ 4.2), its configuration is well matched to the typical image
quality obtained at the Dark Sky New Mexico observatory site.

\begin{figure}[ht!]
\centering
\includegraphics[width=0.45\linewidth]{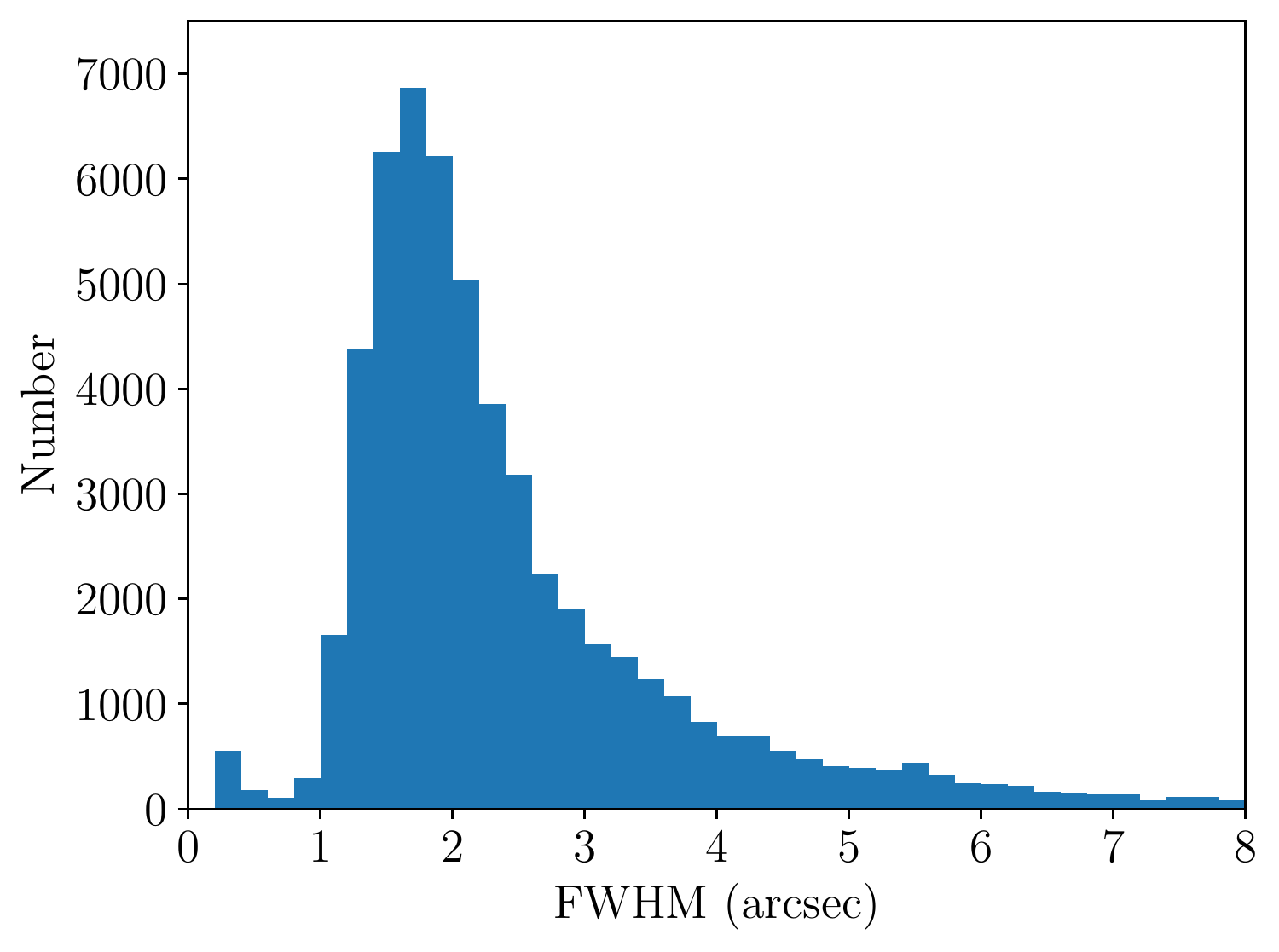}
\caption{Distribution of ``image quality'' FWHM (including effects of seeing,
focus, and tracking errors) determined from measurements of the autocorrelation
function of images obtained through the luminance filter over the time interval
stretching from April 21, 2021 through June 19, 2022.  The mode of the
distribution is 1.6 arcsec, and the median of the distribution is 2.0 arcsec.
Bin width is 0.2 arcsec.}
\end{figure}

\subsection{Night-Sky Background}

  The distributions of the ``night-sky background rate'' (in units e$^-$
s$^{-1}$ pix$^{-1}$) and the ``night-sky surface brightness'' $\mu_{\rm sky}$
(in units mag arcsec$^{-2}$) measured for a large number of images obtained
through the luminance filter are shown in Figure 16.  (The night-sky surface
brightness $\mu_{\rm sky}$ was determined from the night-sky background rate
using a photometric zero point $m_0 = 22.7$ appropriate for clear conditions,
as described in \S\ 4.6.)  In Figure 16a, the large peak results from images
obtained under dark conditions, and the long tail (which continues
substantially beyond the right-hand edge of the plot) results from images
obtained under brighter conditions.  In Figure 16b, the prominent peak near
$\mu_{\rm sky} = 21.5$ mag arcsec$^{-2}$ results from images obtained when the
Moon is below the horizon, and the prominent peak near $\mu_{\rm sky} = 18.5$
mag arcsec$^{-2}$ results from images obtained when the full Moon is above the
horizon.  From Figure 16 it is apparent that under the darkest conditions ever
experienced by Condor, the night-sky background rate is $\approx 0.5$ e$^-$
s$^{-1}$ pix$^{-1}$ and the night-sky surface brightness is $\mu_{\rm sky}
\approx 21.7$ mag arcsec$^{-2}$ through the luminance filter.  The Condor site
is as dark as any ground-based astronomical site, including Muana Kea, of which
we are aware \citep[see Figure 5 of][]{bar2022}.  Measurements of the night-sky
background rate and night-sky surface brightness through other filters are
given in Appendix A.

\begin{figure}[ht!]
\centering
\subfloat{
  \includegraphics[width=0.45\linewidth]{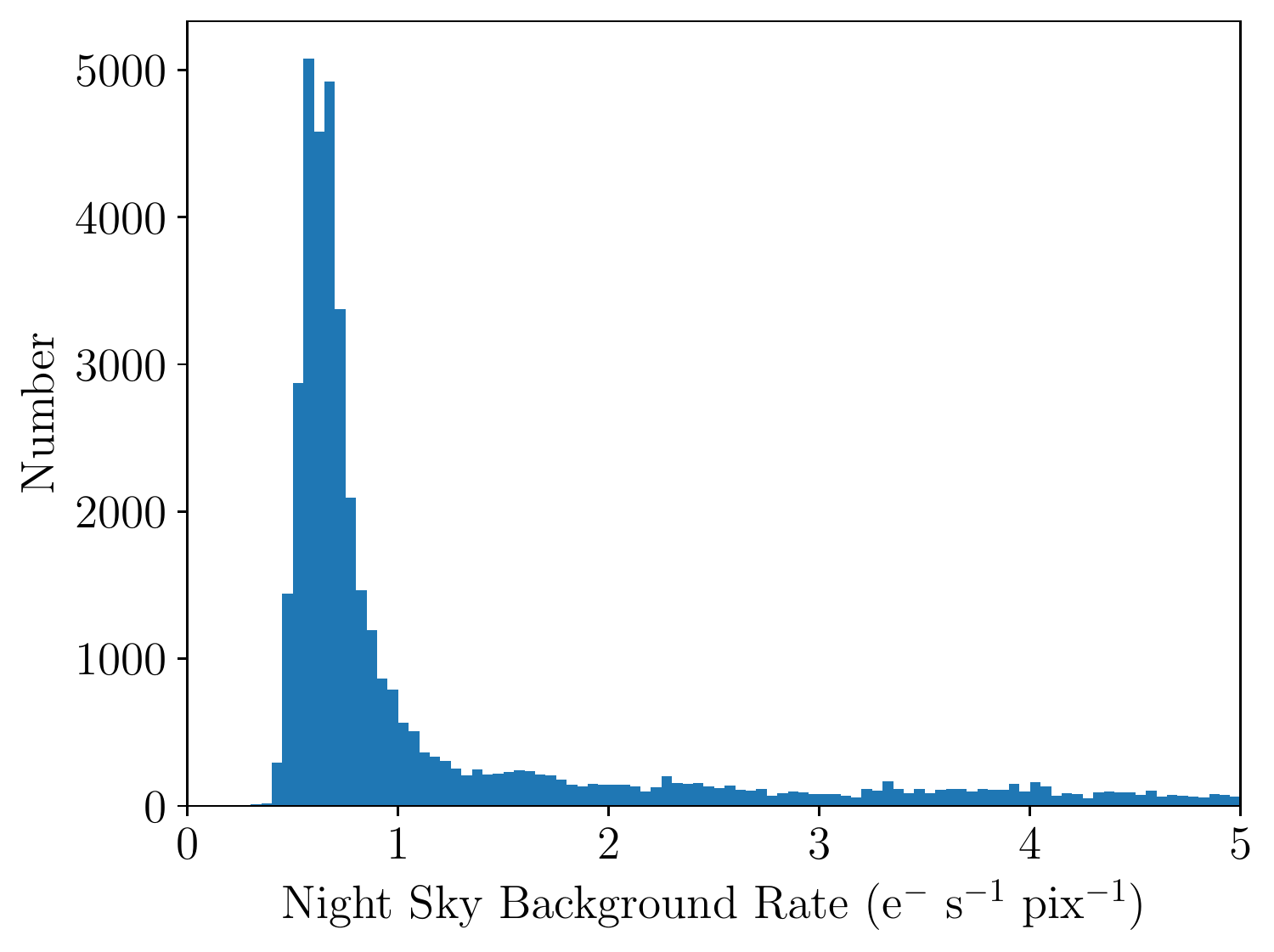}
}
\subfloat{
  \includegraphics[width=0.45\linewidth]{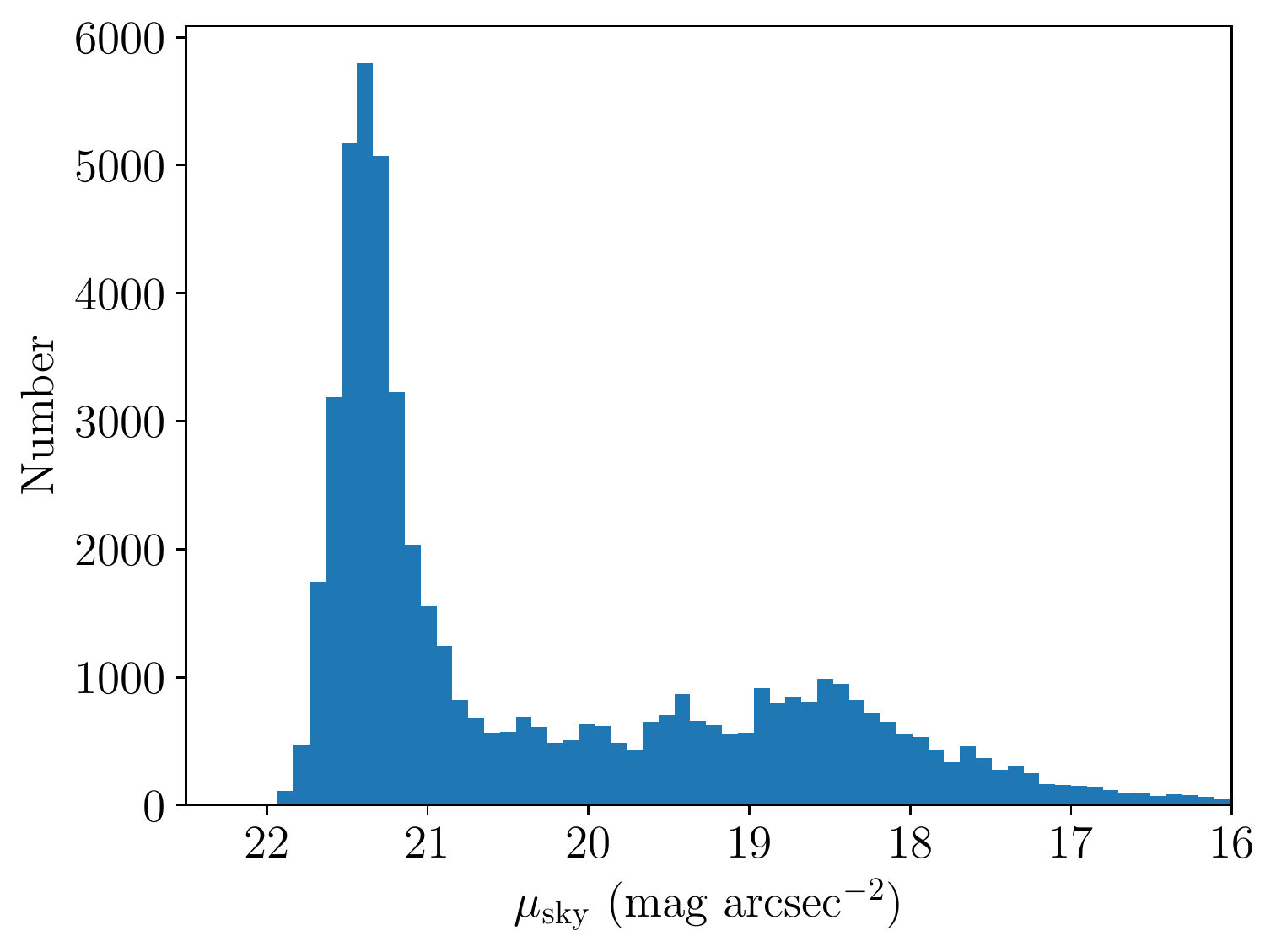}
}
\caption{Distribution of night-sky background of a large number of images
obtained over a time interval stretching from April 21, 2021 through June 19,
2022 through the luminance filter.  The panels show the distribution ({\em a,
left}) in units e$^-$ s$^{-1}$ pix$^{-1}$ and ({\em b, right}) in magnitude
surface brightness units mag arcsec${-2}$, adopting the photometric calibration
described in \S\ 4.6.  Bin width of left panel is 0.05 e$^-$ s$^{-1}$
pix$^{-1}$, and bind width of right panel is 0.065 mag.}
\end{figure}

Over a 60-s exposure, a night-sky background rate of $\approx 0.5$ e$^-$
s$^{-1}$ pix$^{-1}$ produces a sky background of $\approx 30$ e$^-$ per pixel,
which corresponds to a sky noise of $\approx 5.5$ e$^-$ per pixel, which
dominates the read noise 1.2 e$^-$.  We conclude that Condor is sky-noise
limited through the luminance filter in 60-s exposures under the darkest
conditions it ever experiences.  In fact, given the very low read noise of the
CMOS detectors, Condor would be sky-noise limited through the luminance filter
in exposures as short as 20 or even 10 s (which correspond to sky noises of
3.2 or 2.2 e$^-$ per pixel, respectively) under the darkest conditions it ever
experiences.  Hence if called for by the science case, Condor could operate at
a cadence as rapid as 20 or even 10 s through the luminance filter while
remaining sky-noise dominated.  Condor is easily sky-noise limited through the
luminance filter under brighter conditions.

\subsection{Sensitivity}

The point-source and surface-brightness sensitivities of Condor may be directly
determined from the measurements of the CMOS detector read noise described in
\S\ 4.1.2, the CMOS detector gain described in \S\ 4.1.3, the plate scale
described in \S\ 4.2, the photometric calibration and magnitude zero point
described in \S\ 4.6, and the night-sky background rate described in \S\ 4.7.
Various point-source and surface-brightness sensitivities are presented in
Table 4 as functions of lunar phase.  In Table 4, the point-source
sensitivities are $5 \sigma$ determined for a 60-s exposure summing all six
telescopes, assuming optimally-weighted measurements and a seeing ${\rm FWHM} =
1.2$ arcsec (and so a PSF ${\rm FWHM} = 1.6$ arcsec at $\lambda = 700$ nm, as
described in \S\ 4.2) under dark, grey, and bright conditions.  And in Table 4,
the surface-brightness sensitivities are $3 \sigma$ determined for exposure
times ranging from 1 to 300 h summing all six telescopes averaged over $10
\times 10$ arcsec$^2$ regions of the sky under dark, grey, and bright
conditions.  Measurements of the sensitivity through other filters are given in
Appendix A.

\begin{table}[ht]
\centering
\hspace{-0.95in}
\begin{tabular}{p{1.0in}ccp{0.35in}p{0.35in}p{0.35in}p{0.35in}p{0.35in}p{0.35in}}
\multicolumn{9}{c}{{\bf Table 4:}  Sensitivity Through Luminance Filter} \\
\hline
\hline
\multicolumn{1}{c}{} & Point & & \multicolumn{6}{c}{Surface Brightness} \\
\multicolumn{1}{c}{} & Source & & \multicolumn{6}{c}{Over $10 \times 10$ arcsec$^2$} \\
\multicolumn{1}{c}{} & (mag) & & \multicolumn{6}{c}{(mag arcsec$^{-2}$)} \\
\cline{2-2}
\cline{4-9}
\multicolumn{1}{c}{Lunar Phase} & 60 s & & \hfill 1 h & \hfill 3 h & \hfill 10 h & \hfill 30 h & \hfill 100 h & \hfill 300 h \\
\hline
dark \dotfill   & 21.0 & & \hfill 28.5 & \hfill 29.1 & \hfill 29.8 & \hfill 30.4 & \hfill 31.0 & \hfill 31.6 \\
grey \dotfill   & 20.4 & & \hfill 27.8 & \hfill 28.4 & \hfill 29.0 & \hfill 29.6 & \hfill 30.3 & \hfill 30.9 \\
bright \dotfill & 19.7 & & \hfill 27.0 & \hfill 27.6 & \hfill 28.3 & \hfill 28.9 & \hfill 29.5 & \hfill 30.1 \\
\hline
\end{tabular}
\end{table}

\vspace{-0.2in}

\section{Mode of Operation}

In its normal mode of broad-band operation, Condor obtains exposures of
exposure time 60 s over dwell times spanning dozens or hundreds of hours.  In
this way, Condor builds up deep, sensitive images while simultaneously
monitoring tens or hundreds of thousands of point sources per field at a
cadence of 60 s.  In its normal mode of narrow-band operation, Condor obtains
exposures of exposure time 600 s.  In both cases, random dithers of $\approx
15$ arcmin are applied between exposures.  Calibration observations are
typically interspersed between science observations throughout the course of
each night, and images of the dusk and dawn twilight sky are typically obtained
every night.

And in its normal mode of operation, Condor operates completely autonomously,
controlled only by its control and acquisition computers.

\section{Summary and Conclusions}

The ``Condor Array Telescope'' or ``Condor'' is a high-performance ``array
telescope.''  The telescope is comprised of six TEC 180 mm-diameter $f/7$
refracting telescopes, each equipped with (1) an A-P 0.72x QUADTCC-TEC180
four-element telecompressor, (2) A ZWO ASI6200MM monochrome CMOS camera, (3) a
ZWO EFW 7/2'' seven-position filter wheel, (4) a variety of broad- and
narrow-band filters and a diffraction grating, (5) an Optec TCF-Leo low-profile
motorized focuser, and (6) and Optec Alnitak motorized remove dust cover.  The
six telescopes are mounted onto a Planewave L--600 half-for mount with
equatorial wedge.  Condor is located at a very dark astronomical site in the
southwest corner of New Mexico, at the Dark Sky New Mexico observatory.  Condor
enjoys a wide field of view ($2.29 \times 1.53$ deg$^2$ or 3.50 deg$^2$), is
optimized for measuring {\em both} point sources {\em and} extended, very
low-surface-brightness features, and for broad-band images can operate at a
cadence of 60 s (or even less) while remaining sky-noise limited with a duty
cycle near 100\%.  In its normal mode of operation, Condor obtains broad-band
exposures of exposure time 60 s over dwell times spanning dozens or hundreds of
hours, thereby building up deep, sensitive images while simultaneously
monitoring tens or hundreds of thousands of point sources per field at a
cadence of 60 s.  Given its unique capabilities, Condor can access regions of
``astronomical discovery space'' that have never before been studied.

\begin{acknowledgments}
We are particularly grateful to Robin Root for providing invaluable assistance
and for patiently enduring a planned two-week road trip to New Mexico that
turned into a nearly five-month extended stay in New Mexico during the height
of the Covid-19 pandemic, which was crucial to the success of the project.  We
are also particularly grateful to Yuri Petrunin for carefully and meticulously
fabricating six superb instruments.  We are also grateful to many others who
contributed significantly to the project, including
Behzad Barzideh,
Rich Bersak,
Mike Benedetto,
Eric Chen,
Scott Chrislip,
Roland Christen,
Kelly Clayton,
David Cyrille,
John Dey,
Jeff Dickerman,
Lee Dickerman,
Tina Dickerman,
Nancy Dipol,
Lisette Garcia,
Susan Gasparo,
John Green,
Vicki Grove,
Diana Hensley,
Michael Hensley,
Ciara Lanzetta,
Hayley Levine,
Barrett Martin,
Doreen Nicholas,
Daniel Oszust,
Celeste Radgowski,
Dennis Recla,
Teri Sentowski,
Sajesh Singh,
Michael Smith,
Dick Stewart,
Zachary Stone,
Jonathan Tekverk,
Andrew Teresky,
Tobias Weiss,
George Whitney,
and
Dave Zurek.
This material is based upon work supported by the National Science Foundation
under Grants 1910001, 2107954, and 2108234.
\end{acknowledgments}


\software{
  astroalign \citep{ber2020},
  astropy \citep{2013A&A...558A..33A,2018AJ....156..123A},
  django \citep{django},
  Docker \citep{mer2014a},
  DrizzlePac \citep{gon2012},
  numba \citep{lam2015},
  numpy \citep{har2020},
  photutils \citep{bra2020},
  scipy \citep{vir2020},
  SExtractor \citep{1996A&AS..117..393B}
}

\newpage

\appendix

\vspace{-0.2 in}

\section{Photometric Calibration, Night-Sky Background, and Sensitivity through
All Filters}

Most of the observations that have been carried out by Condor so far have been
obtained through the luminance filter.  But some of the observations have been
obtained through the other filters described in \S\ 3.5, particularly the
narrow-band filters.  Here we present the photometric calibration, night-sky
background, and sensitivity of all of the filters described in \S\ 3.5.
Because fewer observations have been obtained through filters other than the
luminance filters, the various distributions formed from observations through
these other filters are subject to larger statistical uncertainties than the
corresponding distributions formed from observations through the luminance
filter.  Accordingly, we consider the values derived here to be representative
rather than definitive.  These values are nevertheless useful in establishing
and documenting the capabilities of Condor.

We determined photometric calibrations and magnitude zero points through all
filters using methods similar to those described in \S\ 4.6.  We determined
night-sky background rate and night-sky surface brightness of all filters using
methods similar to those describe in \S\ 4.8.  And we determined sensitivities
through all filters using methods similar to those described in \S\ 4.9.  The
photometric calibrations, magnitude zero points, night-sky background rates,
and night-sky surface brightnesses of all filters are presented in Table 5, and
the sensitivities of all filters are presented in Table 6.

\begin{table}[ht]
\centering
\hspace{-0.70in}
\begin{tabular}{p{1.5in}ccccc}
\multicolumn{6}{c}{{\bf Table 5:}  Photometric Calibration and Night-Sky
Background Through All Filters} \\
\hline
\hline
\multicolumn{1}{c}{} & & & & Night-Sky & \\
\multicolumn{1}{c}{} & & Photometric & & Background & \\
\multicolumn{1}{c}{} & & Calibration & & Rate & $\mu_{\rm sky}$ \\
\multicolumn{1}{c}{Filter} & Telescope & ($\mu$Jy ADU$^{-1}$ s) & $m_0$ &
(e$^-$ s$^{-1}$ pix$^{-1}$) & (mag arcsec$^{-2}$) \\
\hline
luminance \dotfill          & 4 & 3.5 & 22.7 & 0.5  & 21.7 \\
Sloan $g'$ \dotfill         & 0 & 11  & 21.3 & 0.5  & 21.5 \\
Sloan $r'$ \dotfill         & 2 & 9.5 & 21.6 & 0.4  & 20.9 \\
Sloan $i'$ \dotfill         & 4 & 15  & 21.0 & 0.3  & 21.0 \\
He II 468.6 nm \dotfill     & 0 & 250 & 18.0 & 0.02 & 21.4 \\
{[O III]} 500.7 nm \dotfill & 1 & 300 & 17.8 & 0.02 & 21.3 \\
He I 587.5 nm \dotfill      & 2 & 250 & 18.1 & 0.02 & 20.8 \\
H$\alpha$ 656.3 nm \dotfill & 3 & 450 & 17.4 & 0.02 & 20.8 \\
{[N II]} 658.4 nm \dotfill  & 4 & 400 & 17.4 & 0.02 & 21.0 \\
{[S II]} 671.6 nm \dotfill  & 5 & 450 & 17.4 & 0.02 & 21.1 \\
\hline
\end{tabular}
\begin{flushleft}
\tablecomments{Here ``telescope'' refers to telescope used to make
measurements.}
\end{flushleft}
\end{table}

\begin{table}[ht]
\centering
\hspace{-0.95in}
\begin{tabular}{p{1.5in}cccp{0.35in}p{0.35in}p{0.35in}p{0.35in}p{0.35in}p{0.35in}}
\multicolumn{10}{c}{{\bf Table 6:}  Sensitivity Through All Filters} \\
\hline
\hline
\multicolumn{1}{c}{} & \multicolumn{1}{c}{} & Point & & \multicolumn{6}{c}{Surface Brightness} \\
\multicolumn{1}{c}{} & \multicolumn{1}{c}{} & Source & & \multicolumn{6}{c}{Over $10 \times 10$ arcsec$^2$} \\
\multicolumn{1}{c}{} & \multicolumn{1}{c}{} & (mag) & & \multicolumn{6}{c}{(mag arcsec$^{-2}$)} \\
\cline{3-3}
\cline{5-10}
\multicolumn{1}{c}{Filter} & Lunar Phase & 60 s & & \hfil 1 h & \hfil 3 h & \hfil 10 h & \hfil 30 h & \hfil 100 h & \hfil 300 h \\
\hline
luminance \dotfill          & dark   & 21.0 & & \hfil 28.5 & \hfil 29.1 & \hfil 29.8 & \hfil 30.4 & \hfil 30.8 & \hfil 31.6 \\
\multicolumn{1}{c}{}        & grey   & 20.4 & & \hfil 27.8 & \hfil 28.4 & \hfil 29.0 & \hfil 29.6 & \hfil 30.3 & \hfil 30.9 \\
\multicolumn{1}{c}{}        & bright & 19.7 & & \hfil 27.0 & \hfil 27.6 & \hfil 28.3 & \hfil 28.9 & \hfil 29.5 & \hfil 30.1 \\
Sloan $g'$ \dotfill         & dark   & 19.6 & & \hfil 27.1 & \hfil 27.7 & \hfil 28.4 & \hfil 29.0 & \hfil 29.6 & \hfil 30.2 \\
\multicolumn{1}{c}{}        & grey   & 19.0 & & \hfil 26.4 & \hfil 27.0 & \hfil 27.6 & \hfil 28.2 & \hfil 28.9 & \hfil 29.5 \\
\multicolumn{1}{c}{}        & bright & 18.3 & & \hfil 25.6 & \hfil 26.2 & \hfil 26.9 & \hfil 27.5 & \hfil 28.1 & \hfil 28.7 \\
Sloan $r'$ \dotfill         & dark   & 20.0 & & \hfil 27.5 & \hfil 28.1 & \hfil 28.8 & \hfil 29.4 & \hfil 30.0 & \hfil 30.6 \\
\multicolumn{1}{c}{}        & grey   & 19.4 & & \hfil 26.8 & \hfil 27.4 & \hfil 28.0 & \hfil 28.6 & \hfil 29.3 & \hfil 29.9 \\
\multicolumn{1}{c}{}        & bright & 18.7 & & \hfil 26.0 & \hfil 26.6 & \hfil 27.3 & \hfil 27.9 & \hfil 28.5 & \hfil 29.1 \\
Sloan $i'$ \dotfill         & dark   & 19.5 & & \hfil 27.1 & \hfil 27.7 & \hfil 28.3 & \hfil 28.9 & \hfil 29.6 & \hfil 30.2 \\
\multicolumn{1}{c}{}        & grey   & 18.9 & & \hfil 26.3 & \hfil 26.9 & \hfil 27.6 & \hfil 28.2 & \hfil 28.8 & \hfil 29.4 \\
\multicolumn{1}{c}{}        & bright & 18.3 & & \hfil 25.6 & \hfil 26.2 & \hfil 26.8 & \hfil 27.4 & \hfil 28.1 & \hfil 28.7 \\
He II 468.6 nm \dotfill     & dark   & 17.2 & & \hfil 25.5 & \hfil 26.1 & \hfil 26.8 & \hfil 27.3 & \hfil 28.0 & \hfil 28.6 \\
\multicolumn{1}{c}{}        & grey   & 16.9 & & \hfil 24.8 & \hfil 25.4 & \hfil 26.0 & \hfil 26.6 & \hfil 27.3 & \hfil 27.9 \\
\multicolumn{1}{c}{}        & bright & 16.5 & & \hfil 24.1 & \hfil 24.7 & \hfil 25.3 & \hfil 25.9 & \hfil 26.6 & \hfil 27.2 \\
{[O III]} 500.7 nm \dotfill & dark   & 17.0 & & \hfil 25.3 & \hfil 25.9 & \hfil 26.6 & \hfil 27.1 & \hfil 27.8 & \hfil 28.4 \\
\multicolumn{1}{c}{}        & grey   & 16.7 & & \hfil 24.6 & \hfil 25.2 & \hfil 25.8 & \hfil 26.4 & \hfil 27.1 & \hfil 27.7 \\
\multicolumn{1}{c}{}        & bright & 16.3 & & \hfil 23.9 & \hfil 24.5 & \hfil 25.1 & \hfil 25.7 & \hfil 26.4 & \hfil 27.0 \\
He I 587.5 nm \dotfill      & dark   & 17.3 & & \hfil 25.6 & \hfil 26.2 & \hfil 26.9 & \hfil 27.4 & \hfil 28.1 & \hfil 28.7 \\
\multicolumn{1}{c}{}        & grey   & 17.0 & & \hfil 24.9 & \hfil 25.5 & \hfil 26.1 & \hfil 26.7 & \hfil 27.4 & \hfil 28.0 \\
\multicolumn{1}{c}{}        & bright & 16.6 & & \hfil 24.2 & \hfil 24.8 & \hfil 25.4 & \hfil 26.0 & \hfil 26.7 & \hfil 27.3 \\
H$\alpha$ 656.3 nm \dotfill & dark   & 16.6 & & \hfil 24.9 & \hfil 25.5 & \hfil 26.2 & \hfil 26.7 & \hfil 27.4 & \hfil 28.0 \\
\multicolumn{1}{c}{}        & grey   & 16.3 & & \hfil 24.2 & \hfil 24.8 & \hfil 25.4 & \hfil 26.0 & \hfil 26.7 & \hfil 27.3 \\
\multicolumn{1}{c}{}        & bright & 15.9 & & \hfil 23.5 & \hfil 24.1 & \hfil 24.7 & \hfil 25.3 & \hfil 26.0 & \hfil 26.6 \\
{[N II]} 658.4 nm \dotfill  & dark   & 16.6 & & \hfil 24.9 & \hfil 25.5 & \hfil 26.2 & \hfil 26.7 & \hfil 27.4 & \hfil 28.0 \\
\multicolumn{1}{c}{}        & grey   & 16.3 & & \hfil 24.2 & \hfil 24.8 & \hfil 25.4 & \hfil 26.0 & \hfil 26.7 & \hfil 27.3 \\
\multicolumn{1}{c}{}        & bright & 15.9 & & \hfil 23.5 & \hfil 24.1 & \hfil 24.7 & \hfil 25.3 & \hfil 26.0 & \hfil 26.6 \\
{[S II]} 671.6 nm \dotfill  & dark   & 16.6 & & \hfil 24.9 & \hfil 25.5 & \hfil 26.2 & \hfil 26.7 & \hfil 27.4 & \hfil 28.0 \\
\multicolumn{1}{c}{}        & grey   & 16.3 & & \hfil 24.2 & \hfil 24.8 & \hfil 25.4 & \hfil 26.0 & \hfil 26.7 & \hfil 27.3 \\
\multicolumn{1}{c}{}        & bright & 15.9 & & \hfil 23.5 & \hfil 24.1 & \hfil 24.7 & \hfil 25.3 & \hfil 26.0 & \hfil 26.6 \\
\hline
\end{tabular}
\begin{flushleft}
\tablecomments{Point-source sensitivities are $5 \sigma$ summing all six
telescopes assuming optimally-weighted measurements and a seeing ${\rm FWHM} =
1.2$ arcsec.  Surface-brightness sensitivities are $3 \sigma$ summing all six
telescopes averaged over $10 \times 10$ arcsec$^2$ regions of the sky.}
\end{flushleft}
\end{table}

\clearpage

\bibliography{manuscript}{}
\bibliographystyle{aasjournal}

\end{document}